\documentclass[11pt, oneside]{article}   	
\usepackage{geometry}                		
\usepackage{amssymb,bm,amsmath,amsfonts}
\usepackage{multicol,lipsum,float}
\usepackage{graphicx}

\usepackage{subcaption}
\title{\textbf{Linear stability of the ITER 15 MA scenario against the alpha fishbone}}
\author{\textbf{G. Brochard$^{1,2}$, R. Dumont$^{1}$, H. L\"utjens$^{2}$, X. Garbet$^{1}$}}
\date{}							
\begin{document}
\maketitle
\begin{center}
$^1$\emph{ CEA, IRFM, F-13108 Saint-Paul-lez-Durance, France} \\ $^2$ \emph{CPHT, CNRS, Ecole Polytechnique, Institut Polytechnique de Paris, Route de Saclay, 91128 PALAISEAU}\\
E-mail : guillaume.brochard@polytechnique.edu
\end{center}
\begin{abstract}
\noindent
The stability of the $n=m=1$ alpha-fishbone kinetic-MHD mode on the ITER 15 MA baseline scenario \cite{ITERB} is analyzed using the nonlinear hybrid Kinetic-MHD code XTOR-K. Quantitative agreement is found between the complex frequencies $\omega + i\gamma$ computed with the linear model in \cite{Brochard2018} and XTOR-K's linear simulations. Identical precessional resonance positions in phase space are also found between the linear model and XTOR-K. Linear hybrid simulations performed with XTOR-K on the ITER 15 MA scenario reveal that this configuration is likely to be unstable against the alpha fishbone mode. The fishbone thresholds for kinetic-MHD equilibria with flat q profiles with on-axis safety factor just below unity lies between $\beta_{\alpha,thres}/\beta_{tot} = 6-10\%$, whereas the expected beta ratio on ITER is $\beta_{\alpha}/\beta_{tot} = 15-20\%$ \cite{ITERB}.
\end{abstract}

\section{Introduction}
The stability of a hot magnetized plasma against macroscopic modes in fusion devices can generally be predicted by the MagnetoHydroDynamics fluid theory (MHD), provided several assumptions are verified. In particular, the characteristic frequencies $\boldsymbol{\Omega}$ of the charged particles are required to be much higher or lower than the MHD frequencies $\omega$. It ensures that no resonant interactions $\omega=\textbf{n}\cdot\boldsymbol{\Omega}$ occur between the modes and the particles, where $\textbf{n}$ is the wave mode number. Otherwise, a hybrid Kinetic-MHD formalism is required to describe these resonant interactions. In a tokamak configuration, the particles' characteristic frequencies are functions of the gyro-frequency $\omega_c=\Omega_1$, the bounce/transit frequency $\omega_b=\Omega_2$ and the precessional frequency $\omega_d =  \Omega_3 - \epsilon_bq(\bar{\psi})\Omega_2$. $\epsilon_b$ is 1 for passing particles and 0 for trapped ones, $q(\bar{\psi})$ is the safety factor taken on the particles reference flux surface $\bar{\psi}$. For thermal ion species with temperature $T\sim20$keV in large devices such as ITER, $\omega_c/2\pi\sim10^7$Hz, $\omega_b/2\pi\sim10^4$Hz and $\omega_d/2\pi\sim10^2-10^3$Hz, while the MHD frequencies range between $10^4-10^5$Hz. In burning plasmas, a significant fraction of hot ions with $T\sim 1$ MeV exist due to fusion reactions and non-inductive heating. Since the precessional and bounce/transit frequencies depend on the particle's energy as $\omega_d\propto E$ and $\omega_b\propto \sqrt{E}$, resonant interactions cannot be discarded when considering the stability of macroscopic modes in these plasmas.\\ 
\\ 
In this paper, the stability of the so-called "fishbone" mode driven by alpha particles is studied on the ITER tokamak, with the nonlinear Kinetic-MHD code XTOR-K \cite{Luetjens2010}\cite{Leblond2011}. The fishbone mode results from the resonant interaction between the $n=m=1$ internal kink mode, and fast particles inside the $q=1$ surface. The main kinetic drive of this instability is brought by the trapped particles, through the precessional resonance $\omega=\omega_d$. This mode was first discovered on the PDX tokamak \cite{McGuire1983} when fast particles were injected mostly perpendicular to the magnetic field with neutral beam injectors. This instability was then reproduced on a wide range of devices \cite{Campbell1988}\cite{Heidbrink1990}\cite{Nave1991}\cite{Mantsinen2000}. During the nonlinear phase of the fishbone instability, resonant particles inside the $q=1$ surface tend on average to give away their kinetic energy to the $n=m=1$ mode, which leads to their transport beyond the $q=1$ surface. This instability is potentially detrimental to the burning plasmas that will be generated in the ITER tokamak. In these plasmas, particles are required to yield their kinetic energy to the thermal species, in order to maintain the plasma at temperatures allowing fusion reactions to arise. The transport time of resonant alpha particles was found in hybrid nonlinear simulations \cite{Fu2006}, to be of order $10^3\tau_A \sim 10^{-3}$s, whereas the thermalization time of alpha particles in ITER is in the range $10^{-1}-1$s. $\tau_A = V_A/R_0$ stands for the Alfv\'en time, where $V_A$ and $R_0$ are respectively the Alfv\'en velocity and the major radius. Alphas transported by the fishbone therefore cannot heat up the core plasma through thermalization. Two questions then need to be addressed to evaluate the impact of the alpha fishbone on the fusion efficiency of the ITER tokamak. 1) Is the alpha fishbone likely to be triggered for ITER relevant plasma parameters ? 2) What is the fraction of alphas transported during several fishbone oscillations ?\\ 
\\
The first question is addressed in this paper, with plasma parameters relevant to the ITER 15 MA baseline scenario \cite{ITERB}. Previous numerical \cite{Fu2006} and analytical \cite{Hu2006} studies were conducted on this scenario. Results from these studies differ, including when they employ similar parameters, leaving open the issue of ITER stability with respect to the alpha fishbone. The study conducted here, on similar parameters, is based on linear simulations performed with XTOR-K. This code solves the nonlinear extended resistive two-fluid MHD equations in toroidal geometry, while advancing self-consistently populations of kinetic particles with a full-f method in six dimensions, through a Lorentz equation. Given the recent implementation of XTOR-K's kinetic module, a preliminary verification work of the code regarding the alpha fishbone is necessary. Since few linear  simulations of this instability were conducted in the literature, a verification against linear theory of XTOR-K was preferred to a benchmarking as in \cite{Konies}. For this purpose, XTOR-K linear simulations of the alpha fishbone are compared to the results obtained from the fishbone linear model developed in \cite{Brochard2018}. This model evaluates the solutions of the fishbone dispersion relation \cite{Chen1984}\cite{Porcelli1994}, for an isotropic distribution of fast particles described by a slowing-down distribution function \cite{Stix1972}.\\ 
\\
This paper is structured as follows. Section 2 describes the nonlinear code XTOR-K and the fishbone linear model. The specificities and the main restrictions of this model are highlighted. The verification of XTOR-K by the linear model is presented in section 3. A Kinetic-MHD equilibrium respecting the assumptions of the linear model is firstly discussed. Then, comparisons are provided regarding the complex frequencies $\omega+i\gamma$ and the phase-space positions of the precessional resonance obtained between XTOR-K and the fishbone linear model. In section 4, the code XTOR-K is used to investigate the linear stability of the ITER 15 MA scenario, on parameters allowing comparisons with \cite{Fu2006}\cite{Hu2006}. Summary and conclusions are presented in section 5.
\section{Description of XTOR-K and of the fishbone linear model}
\subsection{The nonlinear hybrid code XTOR-K}
\subsubsection{XTOR-K fluid equations}
The fluid equations solved by XTOR-K are an extension of XTOR-2F's \cite{Luetjens2010}, that take into account different moments of the kinetic populations distributions function, depending on the physical model considered. In this work, bulk diamagnetic drifts are neglected and only one kinetic population is considered : fusion alphas. \\ \\ In XTOR-K, to preserve as much as possible the numerical scheme used in XTOR-2F, the MHD velocity is kept unchanged
\begin{equation}
\textbf{v} = \textbf{v}_{E\times B} + u_{i,\parallel}\hat{\textbf{b}}
\end{equation}
with $ \textbf{v}_{E\times B}$ the cross field drift velocity, $u_{i,\parallel}$ the parallel velocity of bulk ions and $\hat{\textbf{b}}=\textbf{B}/B$ the direction of the magnetic field. The quasi-neutrality is preserved by imposing $n_e = Z_in_i + \sum_kZ_kn_k$, with $n_{e/i}$ the bulk electron/ion density, $Z_{i,k}$ the charge number of the ion/kinetic species considered, and $n_k$ the density of the kinetic specie $k$. The quasi-neutrality equation implies that the electron density is not a variable in XTOR-K's equations. A drift ordering with respect to the small parameter $\rho^* = \rho_L/a\ll1 $\cite{Hazeltine} is used to expand the bulk species' average velocities. $\rho_L$ stands for the particles' gyroradius, and $a$ the tokamak minor radius. Since only the first order of the drift ordering is retained in XTOR-K, and since diamagnetic drifts are neglected in the fluid part of the model in this paper, for a specie $s$, $\textbf{u}_s = \textbf{u}_{s,\parallel} + \textbf{v}_{E\times B}$. Considering the quasi-neutrality equation and the MHD velocity, the averaged bulk species' velocities can be expressed as
\begin{equation}\label{mean}
\textbf{u}_i = \textbf{v}, \ \ \ \textbf{u}_e = \textbf{v} + \bigg[\bigg(\frac{Z_in_i}{n_e}-1\bigg)u_{i,\parallel} + \bigg(\frac{J^{q}_{kin,\parallel} - J_{\parallel}}{en_e}\bigg)\bigg]\hat{\textbf{b} }
\end{equation}
with $\textbf{J}^{q}_{kin} = \sum_kq_kn_k\textbf{v}_k$ the kinetic charge current, $n_k\textbf{v}_k$ being the first order moment of the kinetic population $k$ and $q_k$ the charge of the kinetic population.\\ 
\\
The fluid equations solved by XTOR-K in this case are then simply the single fluid resistive MHD equations with a kinetic coupling in the perpendicular equation of motion. A pressure coupling is used in the code, with $\textbf{P}_k$ the second moment of the kinetic distribution function $k$. XTOR-K therefore solves the following set of fluid equations
\begin{equation}
\partial_t n_i = - \nabla\cdot(n_i\textbf{v}) +\nabla\cdot D_{\perp}\nabla n_i + S
\end{equation}
\begin{equation}
\partial_t u_{i,\parallel} + \Big[(\textbf{v}\cdot\nabla)\textbf{v}-\nabla \nu\nabla \textbf{v}\Big]_{\parallel} = - \frac{\nabla_{\parallel}p_i}{m_in_i} - \frac{Z_i\nabla_{\parallel}p_e}{m_in_e}
\end{equation}
\begin{equation}
\textbf{E} = -\textbf{v}\times\textbf{B} + \eta\textbf{J}
\end{equation}
\begin{equation}
\partial_t\textbf{B} = - \nabla\times\textbf{E}, \ \ \nabla\times\textbf{B} = \textbf{J}
\end{equation}
\begin{equation}
\rho_i\partial_t\textbf{v}_{\perp} + \Big[\rho_i(\textbf{v}\cdot\nabla)\textbf{v} \Big]_{\perp} + \partial_t\sum_k\textbf{J}_{kin,\perp}^m   = \textbf{J}\times\textbf{B} - \bigg[\nabla p_i + \nabla p_e + \nabla\cdot\textbf{P}_k + \nabla\nu\nabla\textbf{v} \bigg]_{\perp}
\end{equation}
\begin{equation}
\partial_tT_i = - \frac{2}{3}T_i\nabla\cdot\textbf{v} - \textbf{v}\cdot\nabla T_i+ 
\frac{1}{n_i} \left(\nabla .n_i\chi_{\perp}^i\nabla T_i+\nabla_{\parallel}. n_i\chi_{\parallel}^i\nabla_{\parallel} T_i\right)+H_i
\end{equation}
\begin{equation}
\partial_tT_e =- \frac{2}{3}T_e\Big[\nabla\cdot\textbf{v}+\nabla_\parallel u_{e,\parallel}\Big] - \textbf{v}\cdot\nabla T_e-u_{e,\parallel}\nabla_\parallel T_e +
\frac{1}{n_e} \left( \nabla .n_e\chi_{\perp}^e\nabla T_e+\nabla_{\parallel}. n_e\chi_{\parallel}^e\nabla_{\parallel} T_e\right)+H_e
\end{equation}
with $\textbf{E}$ the electric field, $p_{e/i}$ and $T_{e/i}$ the bulk electron/ion pressures and temperatures, respectively. $\eta$ the resistivity and $\nu$ the viscosity. $\textbf{u}_e$ is defined in Eq.(\ref{mean}). The kinetic coupling not only implies the kinetic pressure tensor, but also  the kinetic currents $\textbf{J}^{q}_{kin}$ and $\textbf{J}^{m}_{kin} =\sum_k m_kn_k\textbf{v}_k$. The source terms are given by $S=-\nabla\cdot D_{\perp}\nabla n_{i,0}$, $H_i=-1/n_{i,0} \nabla .n_{i,0}\chi_{\perp}^i\nabla T_{i,0}$, and $H_e=-1/n_{e,0} \nabla .n_{e,0}\chi_{\perp}^e\nabla T_{e,0}$ where the subscript 0 refers to the initial equilibrium profiles. In the present work, only the diagonal terms of the total kinetic pressure are kept. The temperature time evolution equation is solved for both bulk electrons and ions, that can have different initial profiles. XTOR-K's fluid set of equations is solved numerically with an implicit Newton-Krylov scheme. 
\subsubsection{XTOR-K kinetic module}
\begin{figure}[H]
\centering
\includegraphics[scale=0.26]{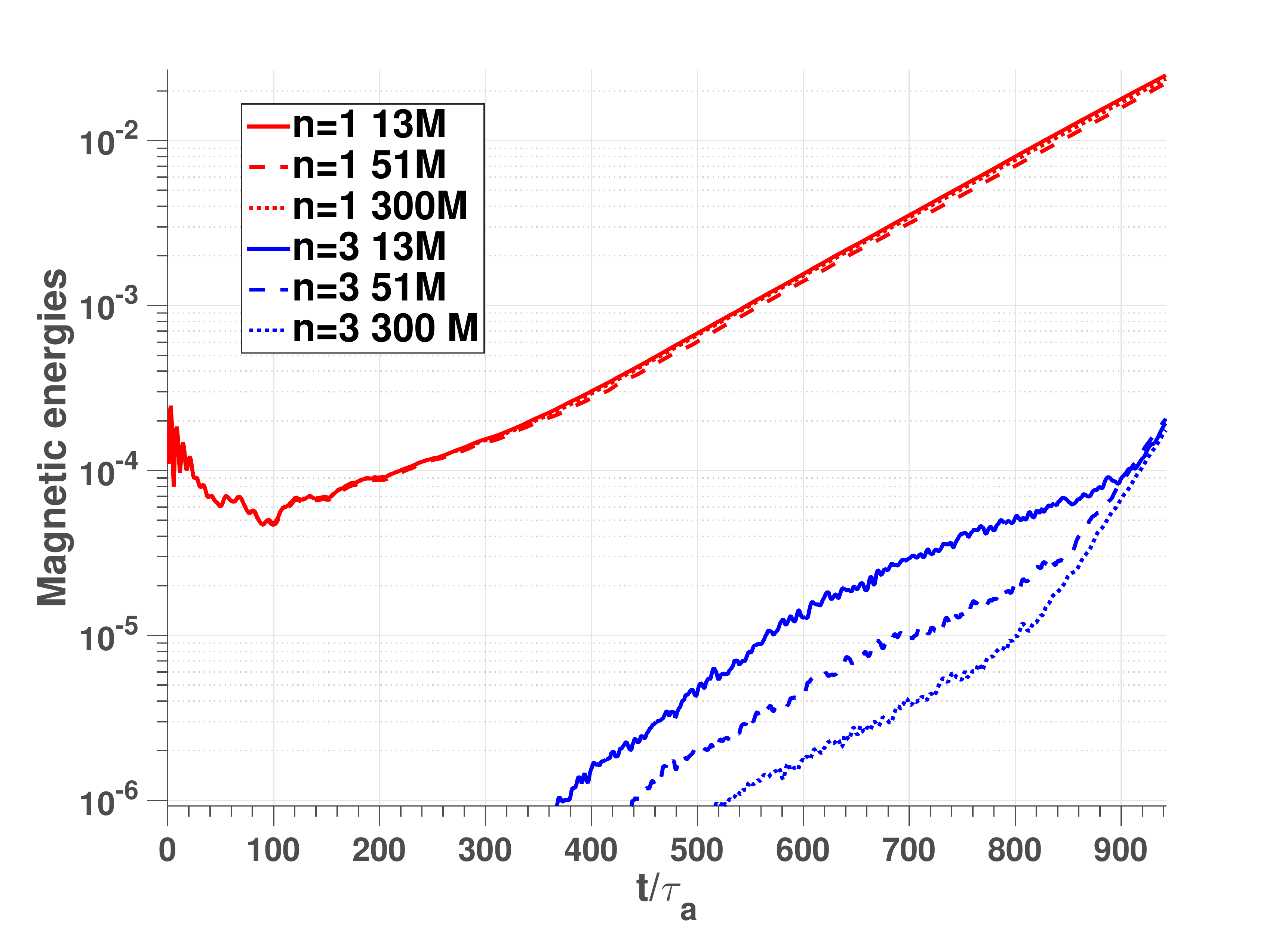}
\caption{Magnetic energies of the $n=1$ and $n=3$ modes of an internal kink simulated with energetic particles. The number of macro-particles have been varied from one simulation to an other.}
\label{noisePIC}
\end{figure}

In XTOR-K, the kinetic module is full-f and 6D, contrarily to other codes which use a gyrokinetic approximation. Therefore, XTOR-K takes into account all kinetic contributions. The kinetic particle distribution function $F_k$ is computed with a Particle In Cell (PIC) module, on an orthogonal direct grid $(R,\varphi,Z)$. A finite number of macro-particles $N$ is used to represent $F_k$. In the code, the weighting is chosen to be the same for all particles. The noise level introduced by a PIC module with constant weighting is $\epsilon_{noise}\propto 1/\sqrt{N}$ \cite{Aydemir1994}.\\ 
\\
Every kinetic particle $(\textbf{r}_{k,n}\textbf{v}_{k,n})$ of the distribution functions is advanced with a Lorentz equation
\begin{equation}
\dot{\textbf{r}}_{k,n} = \textbf{v}_{k,n}, \dot{\textbf{v}}_{k,n} = \frac{q_k}{m_k}[\textbf{E}(\textbf{r}_{k,n}) + \textbf{v}_{k,n}\times\textbf{B}(\textbf{r}_{k,n})]
\end{equation}
This equation is solved numerically with a Boris-Buneman scheme, using a kinetic sub-time step that resolves their gyro-motion.  The electromagnetic field used in the particle advance is taken from XTOR-K's fluid equations. The moments $\textbf{P}_k, \textbf{J}^m_{kin}$ obtained after the particle advance are injected into XTOR-K's fluid equations. This ensures XTOR-K to have a self-consistent scheme.\\ 
\\ 
To assess the noise level in XTOR-K, tests have been conducted with linear simulations of a $n=m=1$ internal kink mode with energetic particles (Figure \ref{noisePIC}). The number of macro-particles in these simulations has been varied from 13 M to 300 M. The simulations are well resolved since the $n=1$ mode is not affected by the number of macro-particles used. For each simulation, the noise level corresponds to the lowest mode energy, in this case $n=3$. Before the $n=3$ mode rises above the noise level due to toroidal pumping with the $n=2$ mode, the magnetic energy ratio between the different simulations indeed varies as $1/\sqrt{N_{phys}}$. 
\subsubsection{Initialization of kinetic particles in XTOR-K}
Realistic distributions of energetic particles have been implemented in XTOR-K to describe energetic ions. The subscript $h$ is now used instead of $k$ to denote energetic ions.
 These distributions can either be isotropic to describe fusion alphas, or anisotropic in the case of NBI or ICRH generated ions. The general distribution function describing these populations is a slowing-down distribution \cite{Stix1972}
\begin{equation}\label{SD}
F_{SD,ani}(r,v,\lambda) = \frac{n_h(r)}{C}\frac{\sigma_H(v_b-v)}{v^3+v_c^3(r)}e^{-\Big[\frac{\lambda-\lambda_0}{\Delta\lambda}\Big]^2}
\end{equation}
with $r$ the radial position, $v$ the velocity norm, $\lambda$ the pitch angle, $v_b$ the birth velocity. The critical velocity $v_c(r)\propto v_{th,e}(r)$ is the velocity at which fast particles yield as much energy $\emph{via}$ thermalization to bulk ions and electrons, with $v_{th,e}$ the electron thermal velocity. When the slowing-down is anisotropic, a Gaussian pitch-angle dependence is assumed, centered around $\lambda_0$ with width $\Delta \lambda$. The slowing down distribution is therefore isotropic when $\Delta \lambda \rightarrow \infty$. $C$ is a normalization factor, and $\sigma_H$ the Heaviside function.\\ 
\\
This distribution is implemented in XTOR-K through random shooting, by inverting the different cumulative probability density functions along each of the phase-space coordinates. The probability density functions along the angles $(\varphi,\theta,\varphi_c)$ are uniform, with $\varphi_c$ the gyroangle in velocity space. These angles are then initialized randomly in $[0,2\pi]$. The probability density functions for the other coordinates are $n_h(r)$ for $r$, and $h(r,v)$ and $g(r,\theta,v)$ for respectively the norm velocity $v$ and the pitch angle $\lambda$
\begin{equation}
h(r,v) = \frac{\sigma_H(v_b-v)}{\ln(1+[v_b/v_c(r)]^3)(v^3+v_c^3(r))}
\end{equation}
\begin{equation}
g(r,\theta,v) = \frac{\exp\Big[-\Big(\frac{\lambda-\lambda_0}{\Delta\lambda}\Big)^2\Big]}{H^*(r,\theta)\sqrt{1-\lambda/H^*(r,\theta)}}
\end{equation}
with $H^*(r,\theta) \equiv B_0/B(r,\theta)$. The denominator of $h$ is related to the normalization constant $C$, whereas that of $g$ is related to the Jacobian of the transformation $(v_x,v_y,v_z)\rightarrow(E,\lambda,\varphi_c)$. $n_h$ and $g$ are inverted numerically, whereas $h$ is inverted analytically. Initially in XTOR-K, kinetic distributions were Maxwellians in velocity space with $T_{h,0} = E_b = m_hv_b^2/2$. With such distributions, the fishbone instability was only triggered at physically unrealistic beta ratio such as $\beta_h/\beta_{tot}\sim30\%-50\%$. The total beta is defined here as $\beta_{tot} = p_{tot}/(B^2/2\mu_0)$, with $p_{tot}$ the total pressure, and the kinetic beta is defined as $\beta_h(n_{h,0}) = p_h(n_{h,0})/(B^2/2\mu_0)$, with $p_h$ defined in Eq. (\ref{pSD}) . As it will be shown later, fishbone modes can be unstable with much lower $\beta_h/\beta_{tot}$ with a slowing-down distribution.
\begin{figure}[H]
\begin{subfigure}{.49\textwidth} 
   \centering
   \includegraphics[scale=0.2]{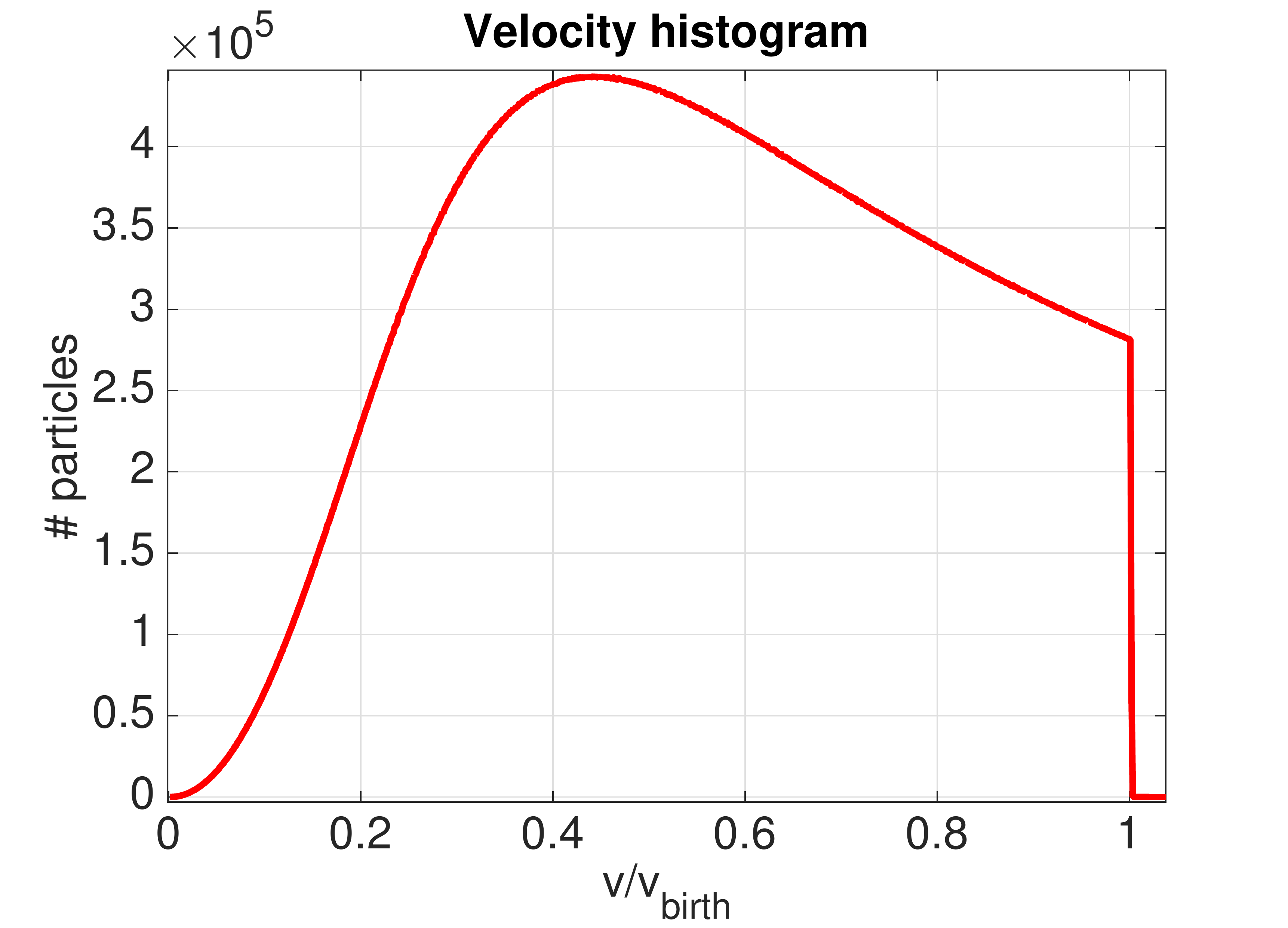}
   \caption{}
\end{subfigure}
\begin{subfigure}{.49\textwidth} 
   \centering
   \includegraphics[scale=0.2]{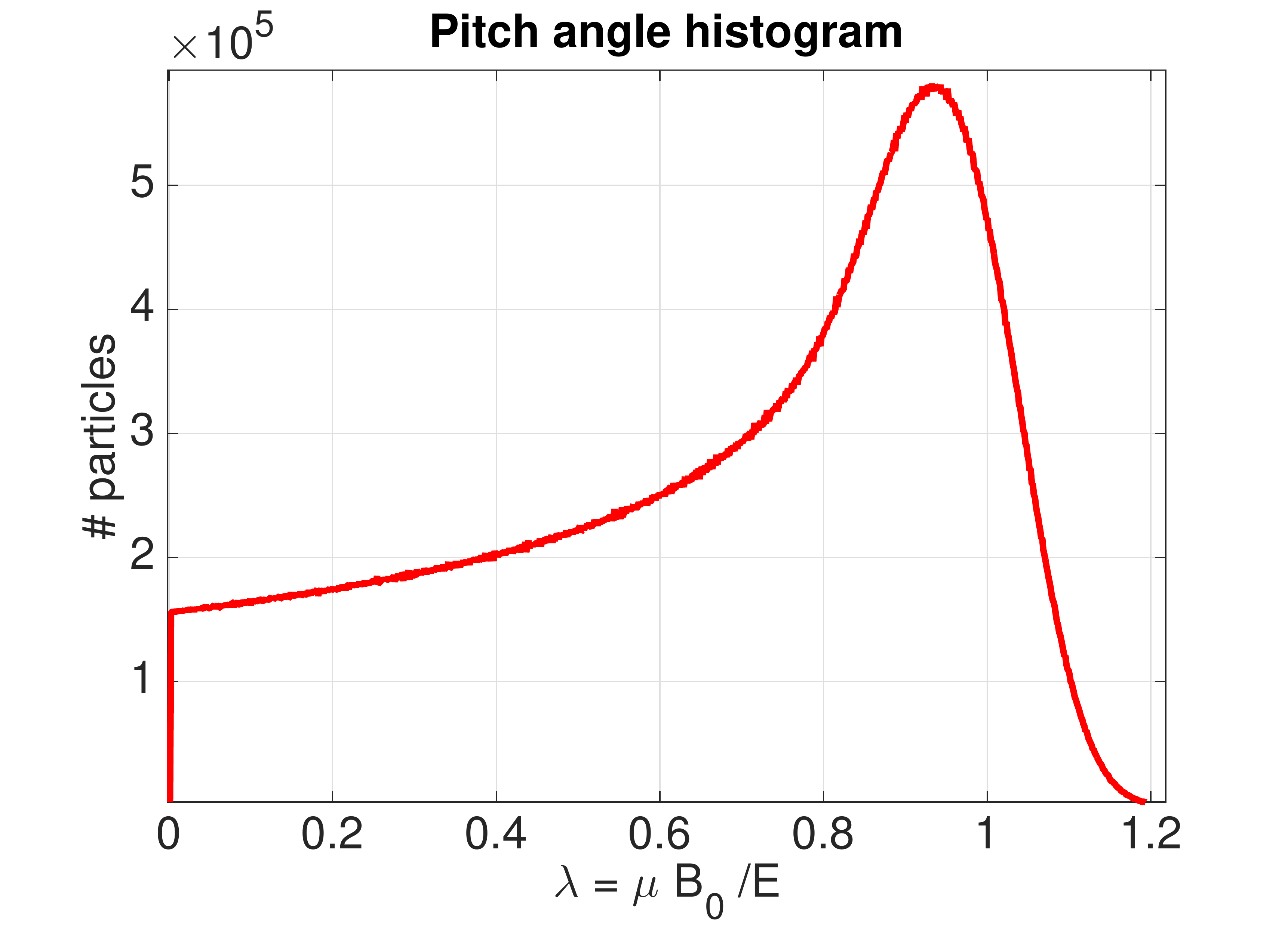}
   \caption{}
\end{subfigure}    
\caption{Number of macro-particles as a function of (a) velocity relative to birth velocity (b) pitch-angle.}
\label{histo}
\end{figure}
Velocity and pitch angle histograms of isotropic slowing down distribution function are illustrated in Figure \ref{histo}, for birth energy $E_{b}=3.5$ MeV and on-axis electron temperature $T_{e,0} = 20$ keV.\\ 
\\
The contribution of kinetic particles is taken into account when solving the Grad-Shafranov equation with CHEASE \cite{Lutjens1996}. The metrics and profiles obtained are therefore coherent with XTOR-K's initialization. Since the kinetic distribution functions are not Maxwellians, the kinetic pressure is computed with
\begin{equation}\label{pSD}
p_h(r) = \int d^3\textbf{v} \ m_hv^2F_h=  n_h(r)E_b\frac{2I_{v2}(r)}{I_{v1}(r)}
\end{equation}
\begin{equation}
I_{v1}= \ln\Big[1+\Big(\frac{v_{b}}{v_c}\Big)^3\Big], I_{v2} = \int_0^1\frac{v^4}{v^3+(v_c/v_b)^3}dv
\end{equation}
The term $E_b^22I_{v2}(r)/I_{v1}(r)$ can be seen as the equivalent kinetic temperature profile when considering Maxwellians distributions with $T_{h,0} = E_b = m_hv_b^2/2$. It is noted that for fusion alphas, the on-axis kinetic pressure computed with a slowing down distribution is smaller by a factor 4 than one computed from such Maxwellian distributions. The CHEASE code only takes into account isotropic total pressure profile. Hybrid simulations of anisotropic distributions require to be started by only letting the harmonic $n=0$ evolve in time, until the profiles have evolved towards a Kinetic-MHD equilibrium coherent with the particles' initialization.
\subsection{The fishbone linear model}
\subsubsection{The fishbone dispersion relation}
The fishbone linear model developed in \cite{Brochard2018} solves non-perturbatively the  fishbone dispersion relation \cite{Chen1984}\cite{Porcelli1994}
\begin{equation}\label{reldisp}
\mathcal{D}(\omega,n_{h,0}) = \omega I_R(\omega) - i\omega_A[\lambda_H + \lambda_K(\omega,n_{h,0})] = 0
\end{equation}
for a given on-axis fast particle density $n_{h,0}$, with $\omega_A = 2\pi/\tau_A$ and $\omega$ is the complex frequency. The term $I_R(\omega)$ is a resistive contribution \cite{Ara1978}, $\lambda_H = \gamma_{MHD}\tau_A$ is a fluid ideal term linked to the MHD part of the mode potential energy in the Energy Principle \cite{Freidberg2014}. In this fishbone model, $\lambda_H$ is computed for a given Kinetic-MHD equilibrium by computing the mode's linear fluid growth rate with XTOR-2F, without the contribution of energetic particles. However, since this term is taking into account the total current $\textbf{J}$ \cite{Bussac1975}, including $\textbf{J}_h$ the current of fast particles, this method can only be accurate when $|\textbf{J}_h|\ll | \textbf{J}|$, $\emph{i.e}$ for low kinetic beta $\beta_h/\beta_{tot} \ll 1$. The fishbone linear model developed here removes this constraint by computing $\lambda_K$ as follows.
\subsubsection{Computation of the $\lambda_K$}
\begin{figure}[H]
\centering
\includegraphics[scale=0.25]{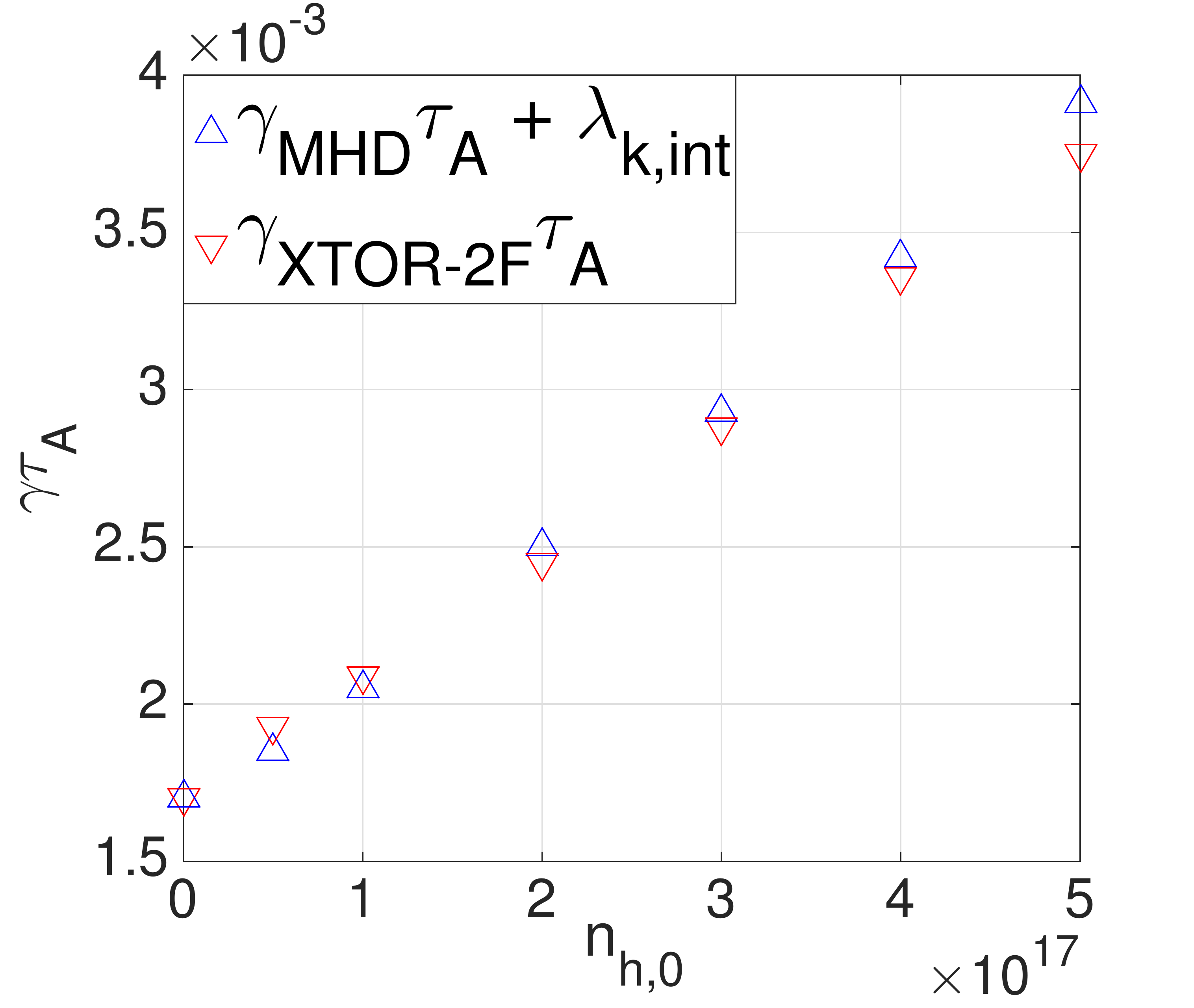}
\caption{Linear growth rates for the internal kink instability as a function of the on-axis kinetic density $n_{h,0}$, obtained from the 
fluid fishbone model and XTOR-2F, for a Kinetic-MHD equilibrium where the resonances have been neglected. Growth rates are normalized to the Alfv\'en time $\tau_A$, and the on-axis density are expressed in m$^{-3}$.}
\label{comp_fluid}
\end{figure}

\begin{figure}[H]
\centering
\includegraphics[scale=0.21]{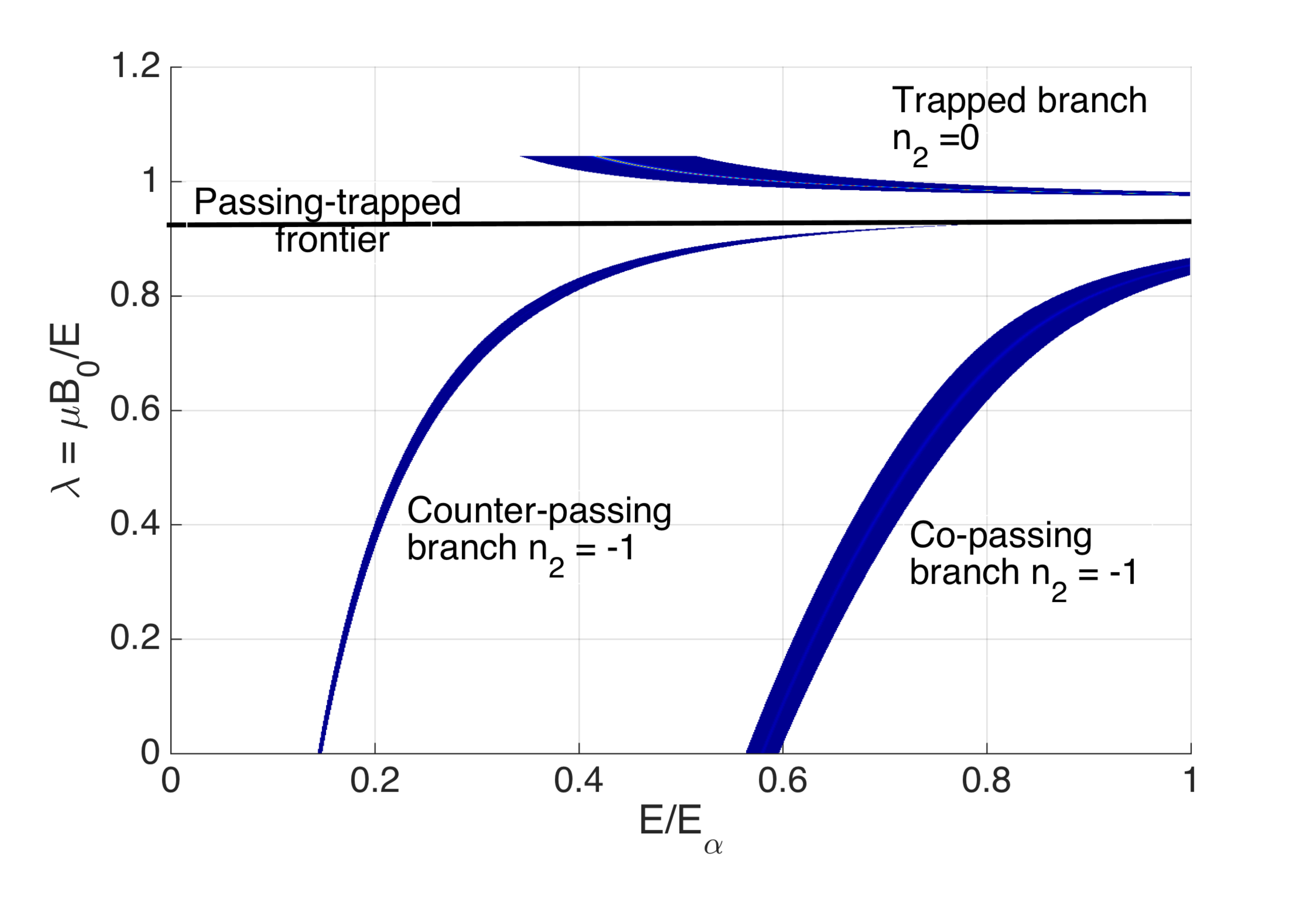}
\caption{Curves of resonances between fast particles and the internal kink in the $(E,\lambda)$ diagram at a fixed $\bar{r}$ position. Three branches exist, the precessional branch with $\textbf{n} = (0,0,1)$ and the co/counter-passing branches with $\textbf{n} = (0,-1,1)$.}
\label{reso_th}
\end{figure}

The term $\lambda_K$ is the kinetic contribution of the mode potential energy, with $\lambda_K \propto \int d^3\textbf{x} \ \boldsymbol{\xi}\cdot\nabla\cdot\tilde{\textbf{P}}_K$, $\boldsymbol{\xi}$ being the MHD displacement and $\tilde{\textbf{P}}_K$ the perturbed kinetic pressure tensor. The MHD displacement is taken at first order such as $\boldsymbol{\xi} = \xi_0 \sigma_H(r-r_{q=1})$, where $r_{q=1}$ is the radial position of the $q=1$ surface. An analytical expression is obtained for $\textbf{P}_K$ by integrating the second order moment of the perturbed kinetic distribution function $\tilde{f}_h$, solution of the linearized Vlasov equation $\partial_t\tilde{f}_h - \{\tilde{h},F_{eq,h}\} - \{H_{eq},\tilde{f}_h\} =0$. $H_{eq}$ and $\tilde{h}$ stand respectively for the equilibrium and perturbed electromagnetic Hamiltonian, $F_{eq,h}$ and $\tilde{f}_h$ for the equilibrium and perturbed distribution functions of hot particles. The analytical expression obtained in \cite{Brochard2018} assumes that $F_{eq,h}$ is an isotropic slowing down as in Eq.(\ref{SD}). This kinetic term can be split into a resonant and a non-resonant part such as $\lambda_K(\omega) = \lambda_K^{int} + \lambda_K^{res}(\omega)$. The interchange term $\lambda_K^{int}$ is the fluid contribution of fast particles. When the resonances between fast particles and the internal kink mode can be neglected, $\lambda_K=\lambda_K^{int}$.  \\ Instead of being computed analytically as in \cite{Brochard2018}, $\lambda_{K,int}$ is computed with XTOR-2F, by adding the pressure profile of kinetic particles computed with Eq.(\ref{pSD}) to the total bulk pressure profile. This is performed by choosing the total bulk ion density profile such $n_{i,tot} = n_i + p_h/T_i$, which enables to form $p_{ion,tot} = n_{i,tot}T_i = n_iT_i + p_h$. The agreement between the two different computations of $\lambda_{K,int}$ is verified in Figure \ref{comp_fluid} where the growth rates obtained from the linear model fishbone and XTOR-2F's linear simulations are compared. \\ 
\\ 
The resonant term $\lambda_K^{res}$ can be expressed as
\begin{equation}\label{kinetic_res}
\lambda_K^{res}(\omega) \propto \sum_{\textbf{n}}\int d^3\textbf{x}d^3\textbf{v}\frac{\partial F_{eq}}{\partial E}\frac{\omega-\omega_*}{\omega-\textbf{n}\cdot\boldsymbol{\Omega}}|Ze(\textbf{v}_d\cdot\nabla)\chi|^2
\end{equation}
with $\omega_*$ is the diamagnetic frequency of the fast particles, $\textbf{v}_d$ the drift velocity, and $\chi=\int_0^tdt'\phi(t')$ the time primitive of the electric potential. In the case of the fishbone instability, fast particles can only interact with the kink through two particle modes, one for trapped particles $\textbf{n} = (0,0,1)$, and one for passing particles $\textbf{n} = (0,-1,1)$. The resonance condition therefore becomes
\begin{equation}\label{resonance}
\omega-\textbf{n}\cdot\boldsymbol{\Omega} = \omega + \epsilon_b\sigma_c[1-q(\bar{r})]\omega_b-\omega_d = 0
\end{equation}
where $\sigma_c$ is the fast particles' parallel velocity sign, and $\bar{r}$ the radial position of their reference magnetic flux surface $\bar{\psi}$. Three branches of resonances result from Eq.$(\ref{resonance})$, the precessional resonance for trapped particles, and the co/counter-passing resonances for passing particles. Since the characteristic frequencies $\omega_d,\omega_b$ depends on the three invariants of motion $(\bar{r},\lambda,E)$, the resonances are planes in 3D invariants space. These resonances are illustrated in Figure $\ref{reso_th}$ in the plane $(E,\lambda)$ at fixed radial position $\bar{r}$. 
The analytical expression of $\omega_d$ depends on the choice made for $\bar{\psi}$ \cite{Graves2013}. In \cite{Brochard2018}, it was chosen to consider $\bar{\psi} = -P_{\varphi}/Ze$, which corresponds to the flux surface of the trapped particles' banana tips. In this paper,  $\bar{\psi} = <\psi>_t$ is used, which corresponds to the time average of particles radial position. This is done to enable a closer comparison between XTOR-K and this linear model. The derivation of $\omega_d$ for an arbitrary $\bar{\psi}$ is detailed in Annex A, and then applied to this convention.\\ 
\\
The kinetic term $\lambda_K$ can be reduced to a triple integral along $(\bar{r},E,\lambda)$ by integrating over the angles $(\theta,\varphi,\varphi_c)$. Only the gyro-center of fast particles is retained (no FLR corrections), and the integral along the poloidal angle is performed by integrating along the energetic particles's poloidal orbit. In order to obtain an analytical expression for $\lambda_K$'s integrand, the MHD equilibria used with the linear model are restricted to shifted circular flux surfaces, obtained from CHEASE. To simplify the analytical calculations, it is also assumed that particles have a thin orbit width. \\ \\ The integral along the invariants $(\bar{r},E,\lambda)$ is performed numerically, using the ''collocation'' method that enables to compute precisely the real and imaginary parts of the resonant integral. This method is presented in Annex B, along with precise expressions for $\lambda_K$'s integrand. These expressions incorporate some corrections compared to \cite{Brochard2018}.
\subsubsection{Specificities and restrictions of the linear model}
The present linear model possesses specificities that are not always taken into account in other linear Kinetic-MHD models \cite{Nabais2015}\cite{Cheng1992}. First of all, the fishbone dispersion relation is solved non-perturbatively, $\emph{i.e.}$ the perturbation of the mode complex frequency $\omega+i\gamma$ due to the kinetic contribution $\lambda_K$ is not considered small compared to $\gamma_{MHD}$. Such a feature is crucial for the fishbone instability, since on the fishbone branch, solution of Eq.(\ref{reldisp}), the mode growth rate can be much larger than $\gamma_{MHD}$. It then prevents to linearize Eq.(\ref{reldisp}) in order to solve it explicitly, as it is the case in \cite{Nabais2015}. Second, the fluid contribution of fast particles $\lambda_{K,int}$ is also taken into account, since the modification of $\beta_{tot}$ imposed by fast particles is non-negligible. Finally, the fluid and resonant contributions of passing particles are taken into account in $\lambda_K$, which is not the case in \cite{Cheng1992}\cite{Fu2006}.
\\ 
\\ 
The fishbone linear model is however limited by a number of restrictions. In order to integrate analytically the particles' poloidal orbit, the considered equilibria need to be concentric circular, and fast particles need to have a thin orbit width. This second requirement limits the kinetic energy of fast particles' in the model. Particles with high kinetic energy have a significant banana/potato width, that are not retained in this simplified model. These two restrictions prevent the fishbone linear model from being directly employed to study the stability of the ITER 15 MA configuration against the alpha fishbone, due to its shaped equilibrium and the high birth energy of fusion alphas.
\section{Linear verification of XTOR-K}
\subsection{Relevant Kinetic-MHD equilibrium for linear verification}
In order to verify XTOR-K with the fishbone linear model, a Kinetic-MHD equilibrium that suits its restricting assumptions is required. The concentric circular flux surfaces requirement is easy to satisfy with CHEASE. However, the thin orbit width approximation requires special care. \\ 
\\ 
The particle energy beyond which the thin orbit width approximation breaks down can be found by comparing values for the particles characteristic frequencies between XTOR-K and the analytical expression for $\omega_b$ and $\omega_d$ developed in the fishbone linear model. Since the particle advance in XTOR-K is not restricted at high kinetic energy, the thin orbit approximation breaks down when a significant mismatch appears with increasing energy between the compared values of $\omega_b,\omega_d$.  \\ 
\\ 
However, first of all, the particles' characteristic frequencies obtained from XTOR-K must be verified analytically, for a given kinetic energy where the thin orbit width approximation is well respected. 
\begin{figure}[H]
\begin{subfigure}{.24\textwidth} 
   \centering
   \includegraphics[scale=0.15]{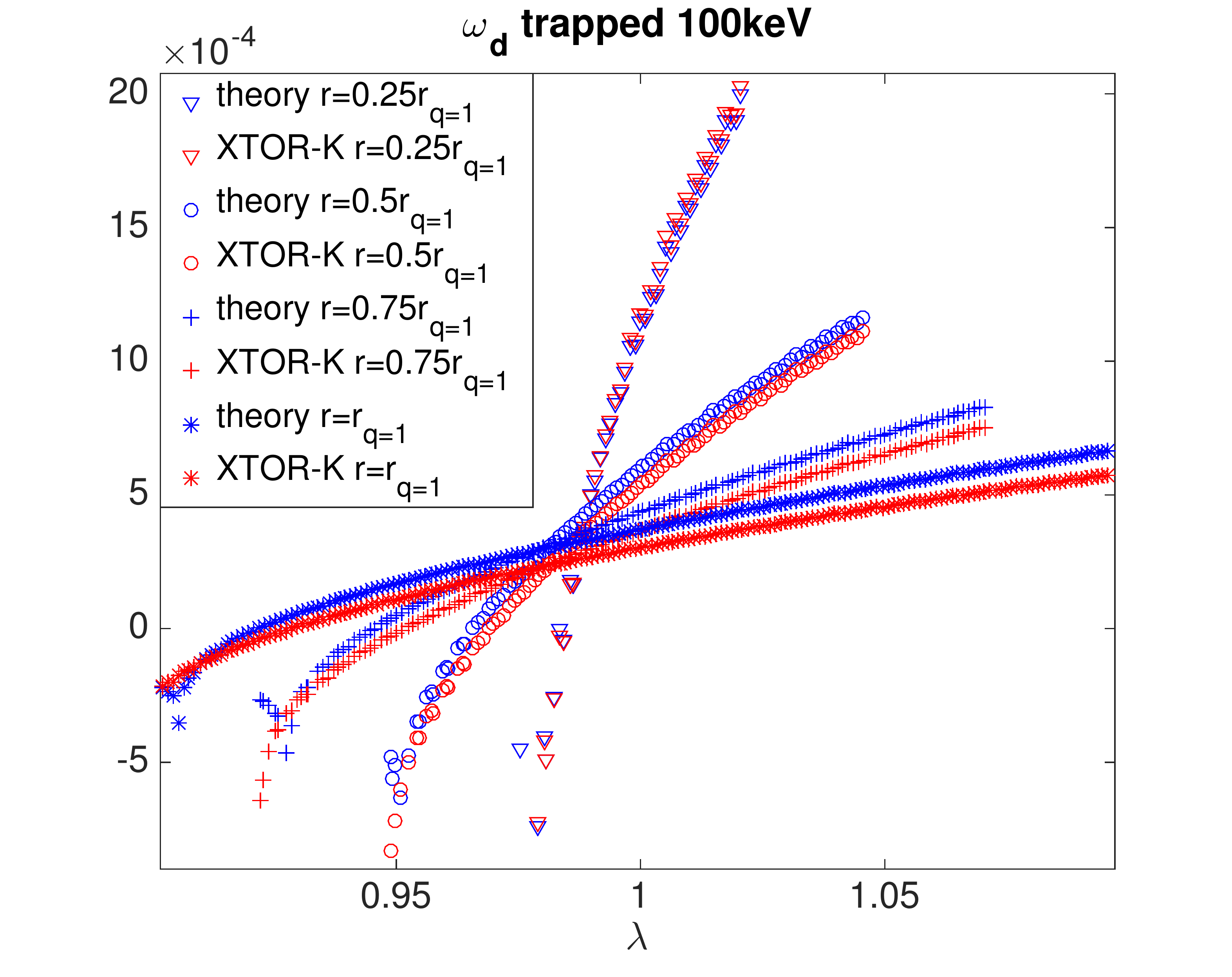}
   \caption{}
\end{subfigure}
\begin{subfigure}{.24\textwidth} 
   \centering
   \includegraphics[scale=0.15]{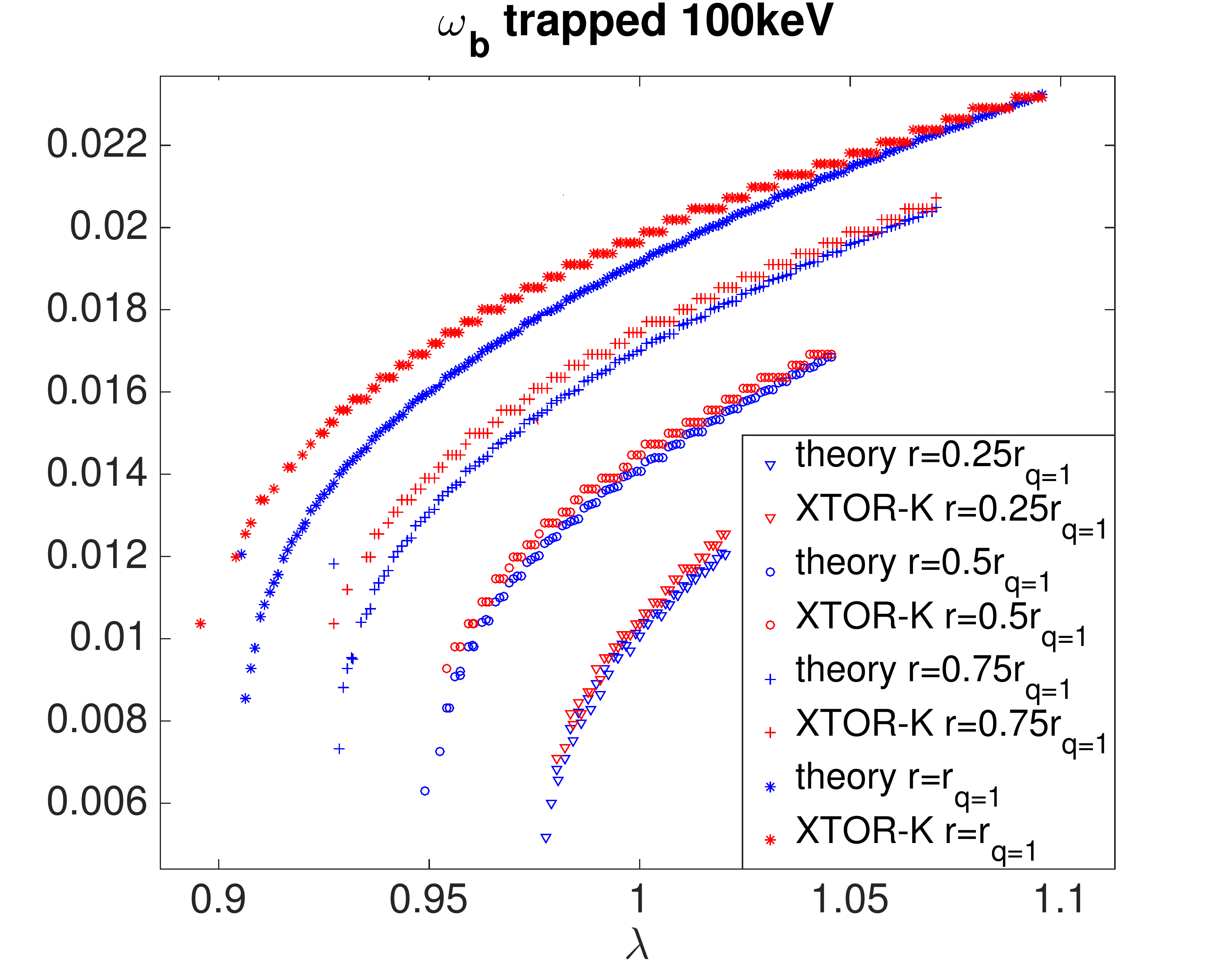}
   \caption{}
\end{subfigure}
\begin{subfigure}{.24\textwidth} 
   \centering
   \includegraphics[scale=0.15]{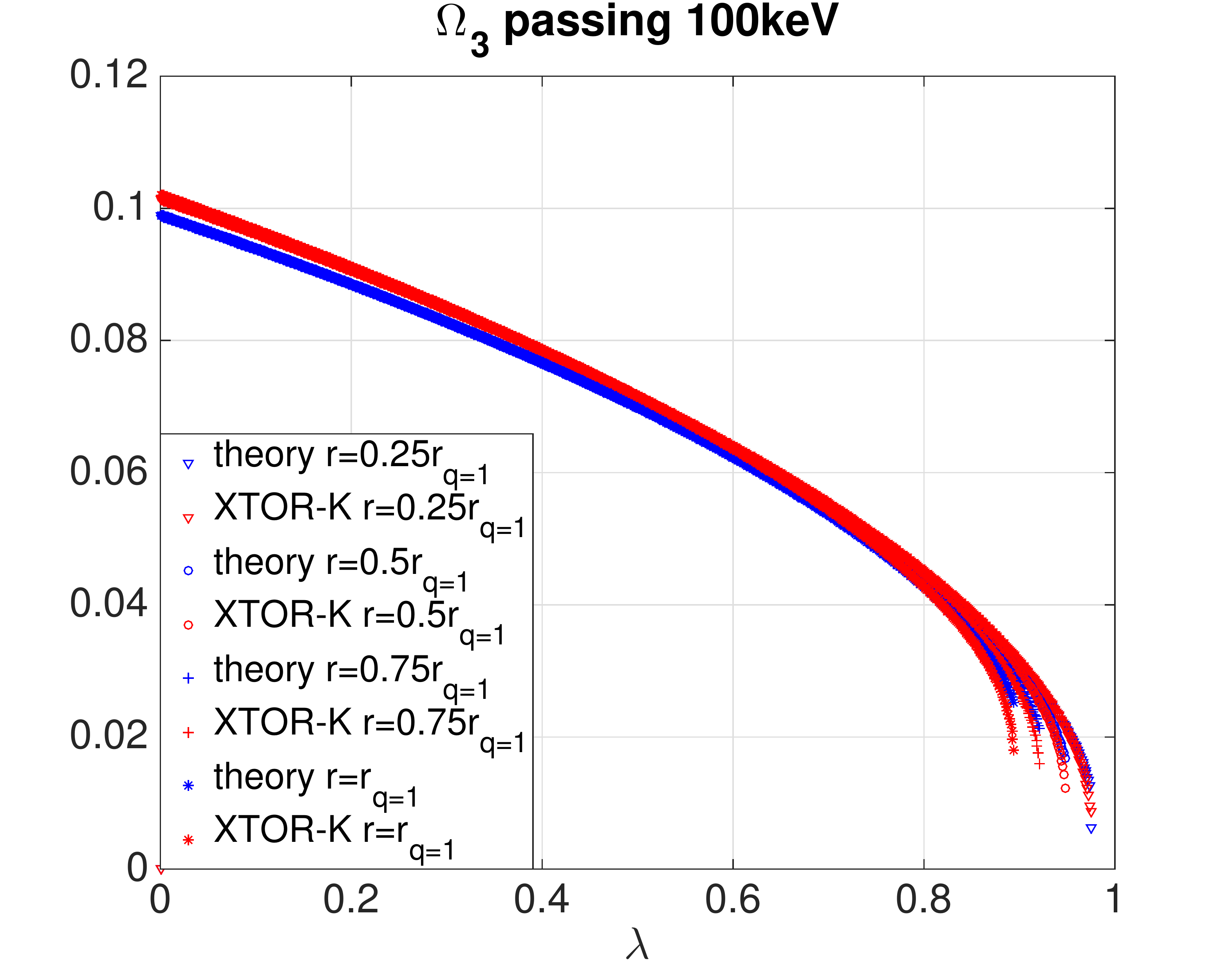}
   \caption{}
\end{subfigure}
\begin{subfigure}{.24\textwidth} 
   \centering
   \includegraphics[scale=0.38]{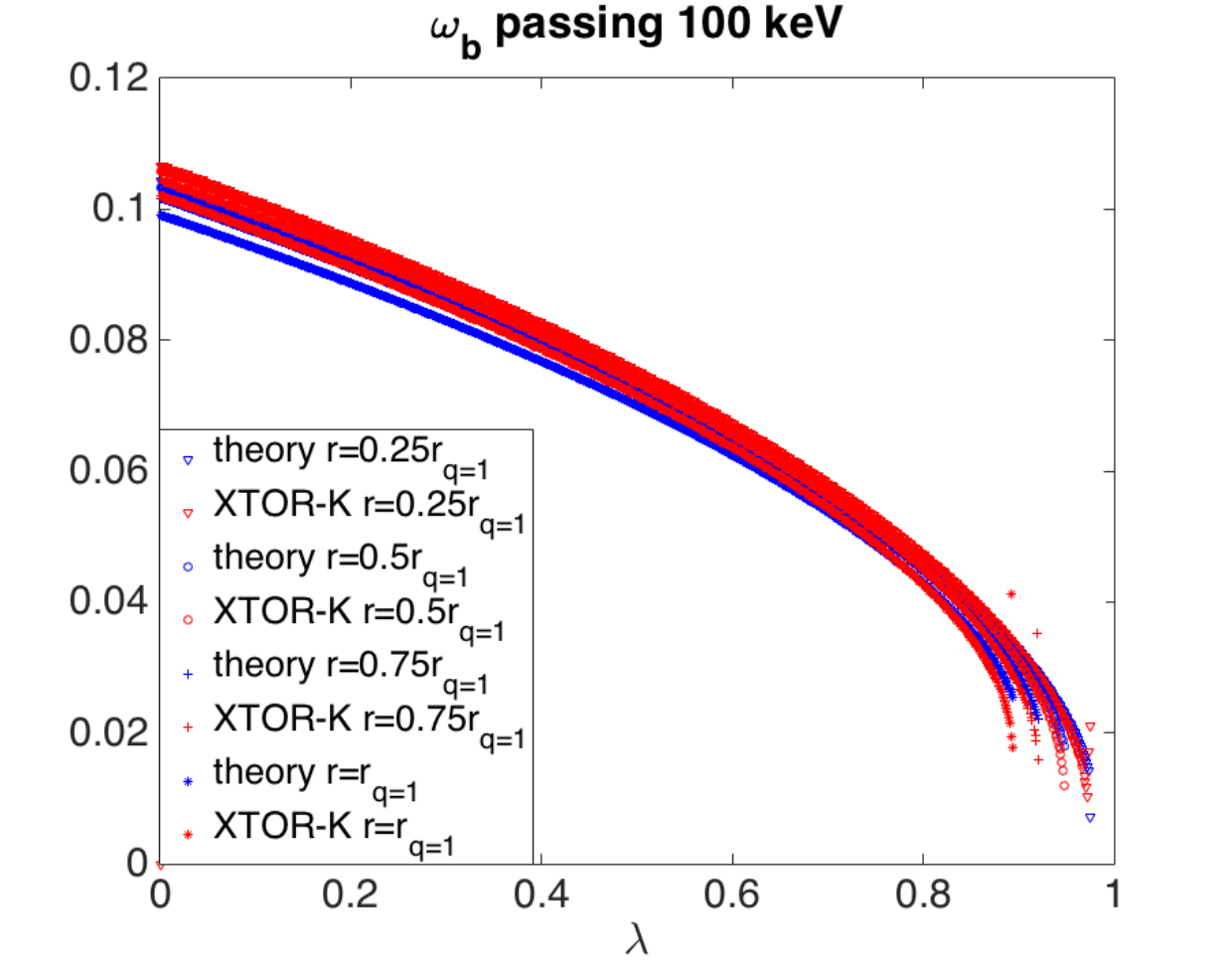}
   \caption{}
\end{subfigure}    
\caption{(a) Precessional frequencies. (b) Bounce frequencies in the trapped domain. (c) Third eigenfrequencies $\Omega_3$. (d) Transit frequencies in the passing domain. Blue points are obtained from the linear model, and the red points are computed from XTOR-K. The frequencies are plotted at different radial positions $\bar{\psi}$ against their pitch angle $\lambda$ for $E=100$ keV. Frequencies are normalized at the Alfv\'en time $\tau_A$.}
\label{freq_check}
\end{figure}
This verification is performed in Figure \ref{freq_check} for particles with kinetic energy $E=100$keV, at different radial positions $\bar{r}$ inside the $q=1$ surface, on a pitch angle range. The trapped precessional frequency $\omega_d$, and the passing third frequency $\Omega_3$ are obtained from a linear regression of the particles toroidal angle $\varphi(t)$. The bounce/transit frequency is computed from a Fourier transform of the particles poloidal angle $\theta(t)$. As seen in Figure \ref{freq_check}, the agreement between the analytical expressions and XTOR-K is satisfactory, for all radial positions considered and all over the pitch angle range. Particles characteristic frequencies are therefore well described by XTOR-K. \\ 
\\
\begin{figure}[H]
\begin{subfigure}{.49\textwidth} 
   \centering
   \includegraphics[scale=0.18]{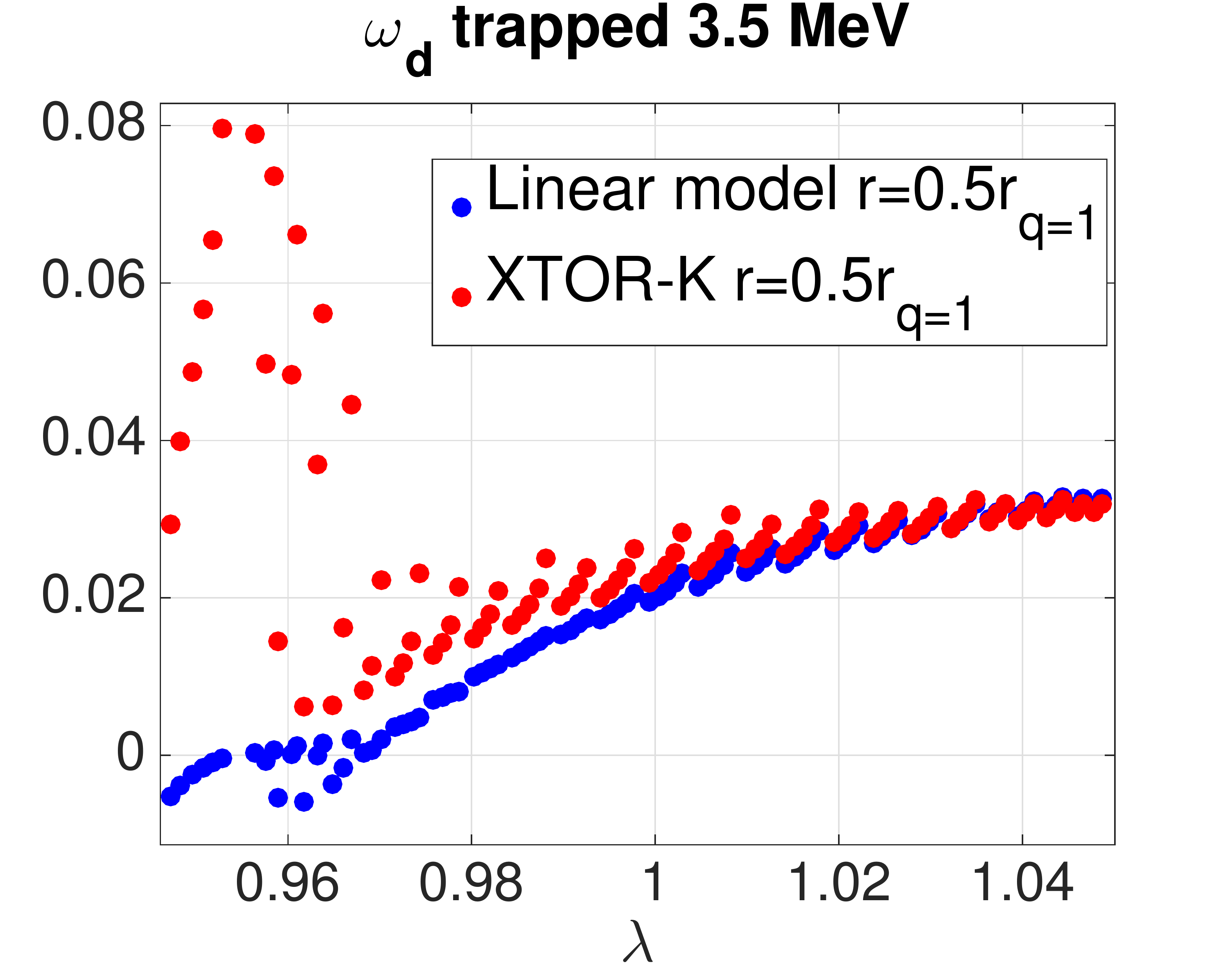}
   \caption{}
\end{subfigure}
\begin{subfigure}{.49\textwidth} 
   \centering
   \includegraphics[scale=0.18]{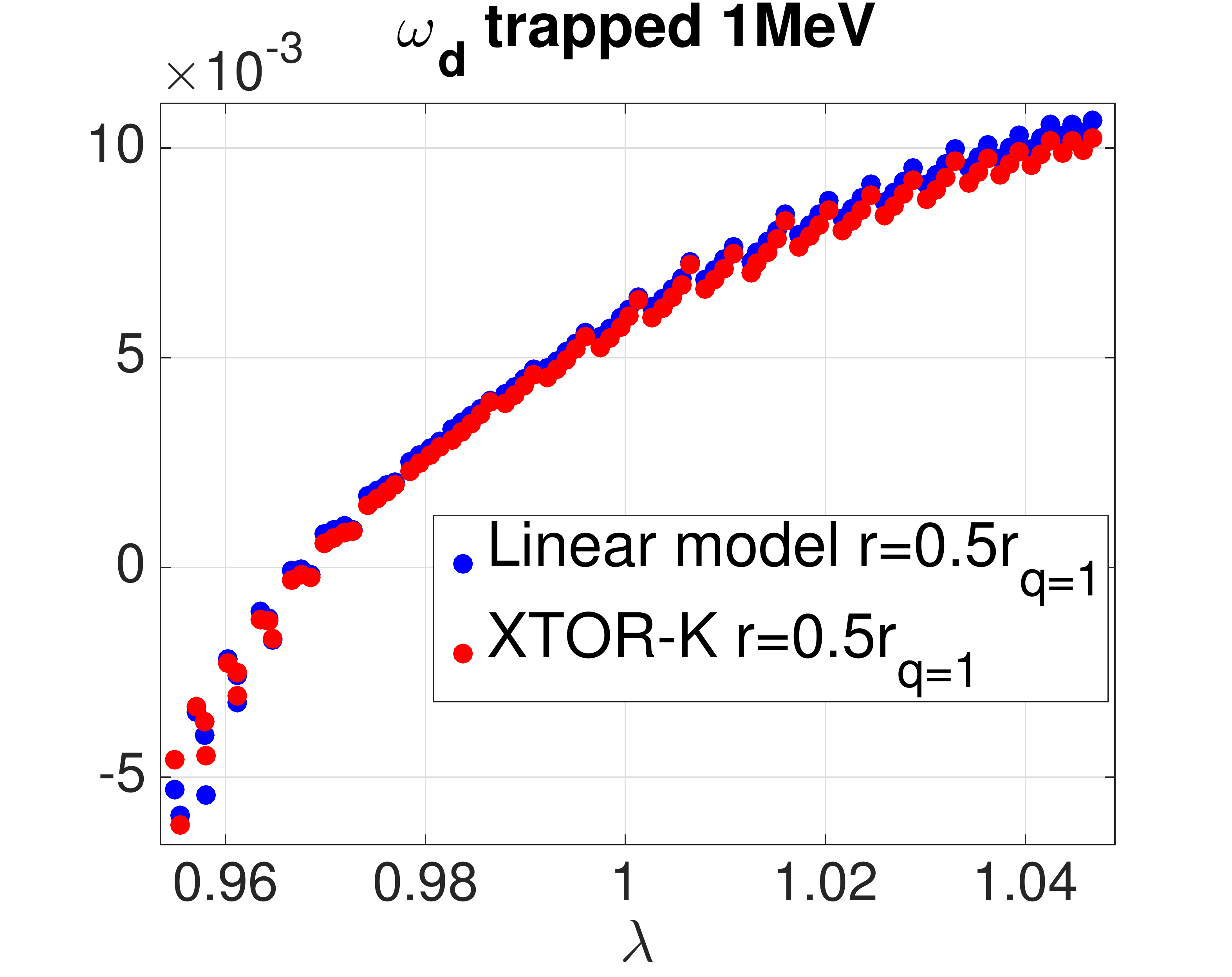}
   \caption{}
\end{subfigure}    
\caption{Comparison of the precessional frequency $\omega_d$ at different energies between the linear model (blue points) and XTOR-K (red points), at r=0.5$r_{q=1}$ against pitch angle in the trapped domain. Results at 3.5 MeV and 1MeV are respectively displayed on (a) and (b).}
\label{freq_verif}
\end{figure}
In Figure \ref{freq_verif}, the precessional frequencies of trapped particles are compared for higher kinetic energies of 1 MeV and 3.5 MeV, for a given radial position. It can be observed that at $E=1$MeV, the agreement between XTOR-K and the analytical expressions is still correct, which means that the thin orbit width approximation holds at this energy. At 3.5 MeV, a significant difference is observed near the trapped-passing boundary at $\lambda=0.94$. When trapped particles are close to the trapped-passing boundary, their orbit are of potato type rather than banana type. This implies that the particles' excursion from their reference flux surface $\bar{\psi}$ is important, giving $\omega_d,\omega_b$ and $\lambda_K$  significant differences between the thin orbit width approximation and full simulations. For this reason, an  isotropic slowing-down distribution functions with birth energy $E_b = $ 1MeV is chosen for the verification of XTOR-K, thus ensuring that the poloidal orbits described are well by the thin orbit width approximation.\\ 
\\
The Kinetic-MHD equilibrium used to verify XTOR-K is an ITER-like circular equilibrium, with $R_0 = 6.2$m, $a=2$m, $B_0 =5.3$ T and Lundquist number $S=1.10^7$. The $q$ profile is parabolic, with on-axis value $q_0=0.95$ and the $q=1$ surface located at $s=0.4$, where $s=\sqrt{\psi/\psi_{edge}}$. The kinetic density profile is chosen rather peaked, with $n_h(s) = n_{h,0}(1-s^2)^6$. 
\subsection{Comparaison of linear results between XTOR-K and the fishbone model}
\subsubsection{Quantitative agreement for the mode complex frequencies}
\begin{figure}[H]
\begin{subfigure}{.5\textwidth}
\centering
\includegraphics[scale=0.25]{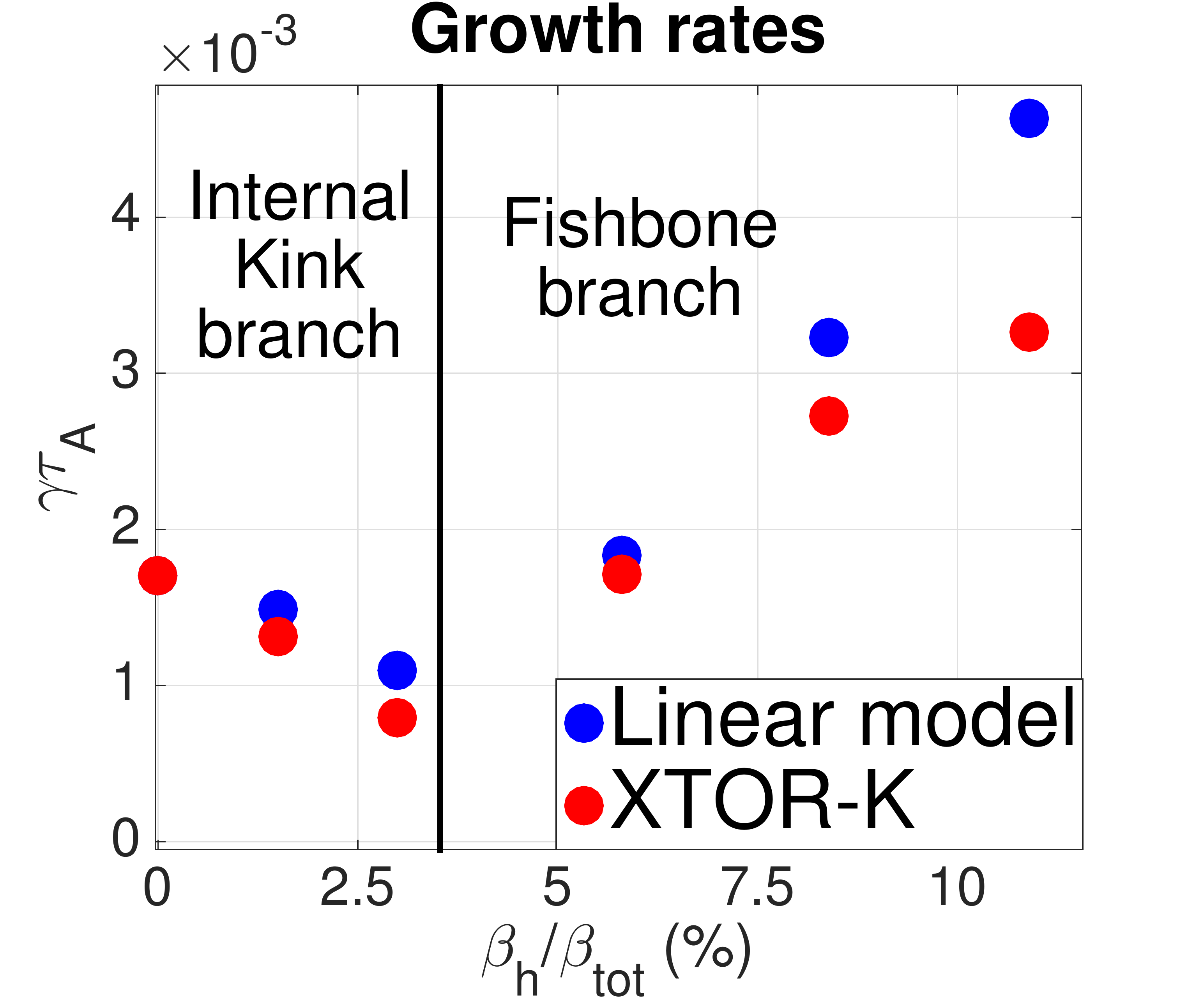}
\caption{}
\end{subfigure}
\begin{subfigure}{.5\textwidth}
\centering
\includegraphics[scale=0.25]{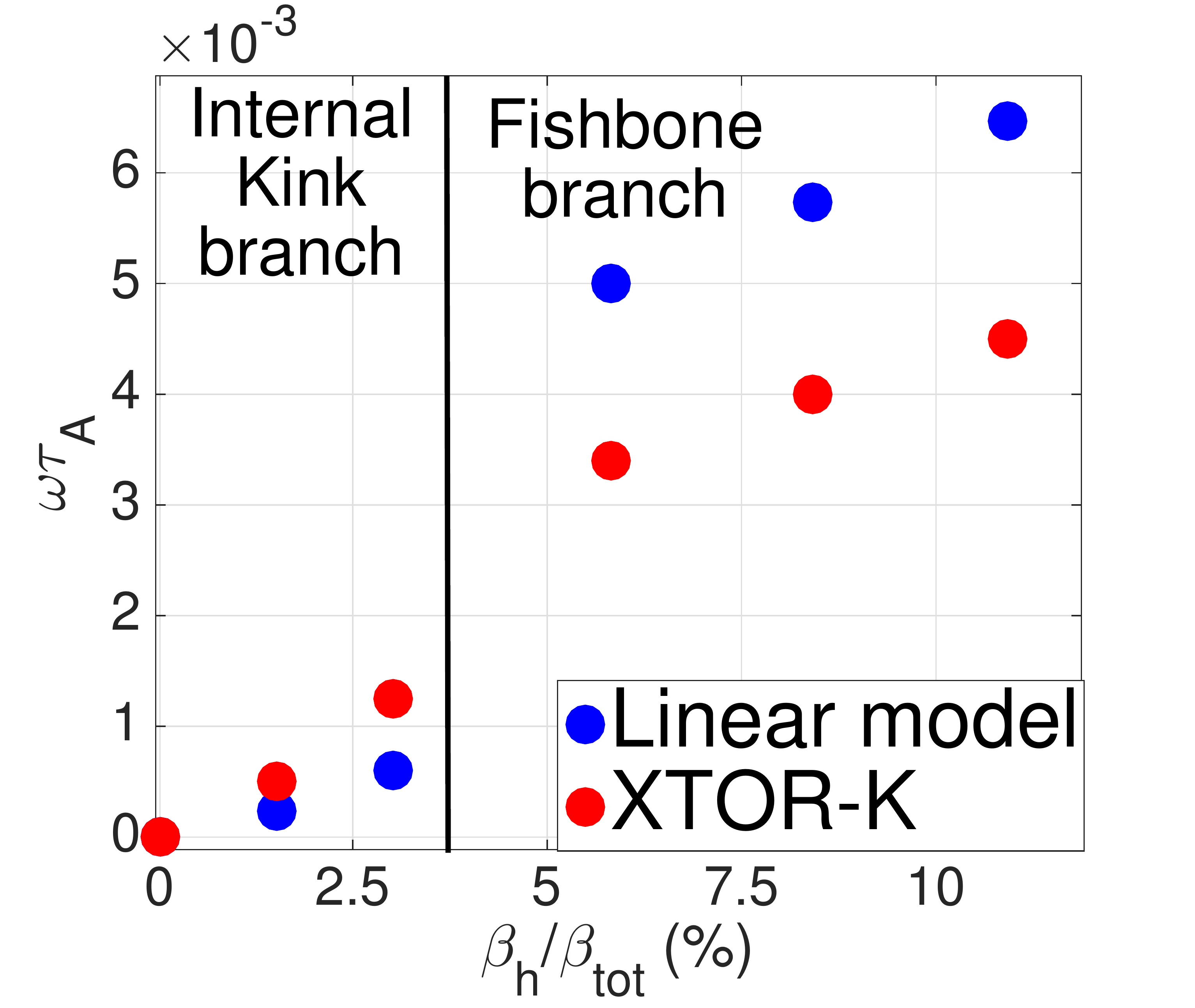}
\caption{}
\end{subfigure}
\caption{Compared complex frequencies against the beta ratio $\beta_{h}/\beta_{tot}$. Values from the linear model in blue, and from XTOR-K in red. Growth rates are shown in (a), frequencies in (b). A quantitative agreement is obtained between the hybrid code and the linear theory.}
\label{Verif}
\end{figure}

The complex frequencies obtained with XTOR-K and the fishbone linear model for the described Kinetic-MHD equilibrium are displayed in Figure \ref{Verif}, as functions of the on-axis kinetic density $n_{h,0}$. In this section and in the rest of the paper, $\omega$ stands for the real part of the mode complex frequency, and $\gamma$ its imaginary part. As expected experimentally and theoretically \cite{Nave1991}\cite{White1990}\cite{Wu1994}, the analytical model and the code XTOR-K recover the internal kink branch and the fishbone branch. At low $n_{h,0}$, the internal kink is stabilized by the kinetic effects of alpha-like particles. At higher $n_{h,0}$ beyond the fishbone threshold $\beta_h/\beta_{tot}=5.5\%$ in Figure \ref{Verif} (a), the fishbone mode is destabilized by the resonant drive, dominating the kink branch. The fishbone mode is therefore an Energetic Particle Mode which only exists in presence of supra-thermal particles.
Regarding the mode's real frequencies, the two branches exhibit different behaviors. The mode frequency on the kink branch is one order of magnitude lower than the one of the fishbone branch. Still on the kink branch, the instability rotates mainly because of the diamagnetic effect carried by the fast alpha-like particles. Instead, on the fishbone branch, the rotation is due to both the diamagnetic effect and the resonant interaction between the $n=m=1$ mode and the fast particles characteristic frequencies. 
On the fishbone branch, due to the resonant interaction, the mode frequency tends to scale as the precessional frequency of deeply trapped particles. In Figure \ref{Verif}, $\omega\tau_A \sim 5.5\times10^{-3}$ and in Figure (\ref{freq_verif}) (b), $ \omega_d\tau_A = 6.10^{-3}$ for $\lambda \sim 1$.  \\ 
\\
\begin{figure}[H]
\begin{subfigure}{.49\textwidth} 
   \centering
   \includegraphics[scale=0.5]{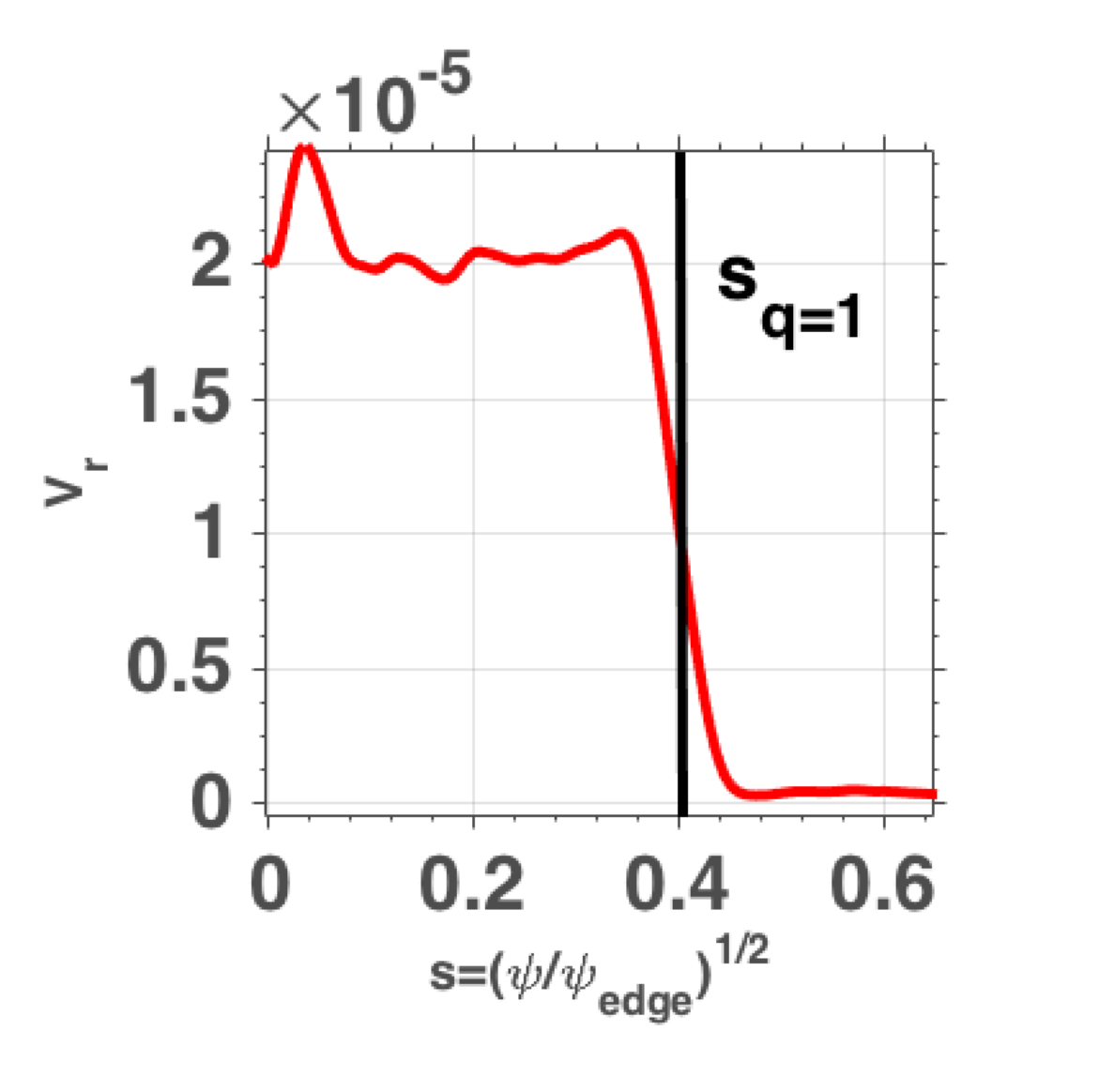}
   \caption{}
\end{subfigure}
\begin{subfigure}{.49\textwidth} 
   \centering
   \includegraphics[scale=0.5]{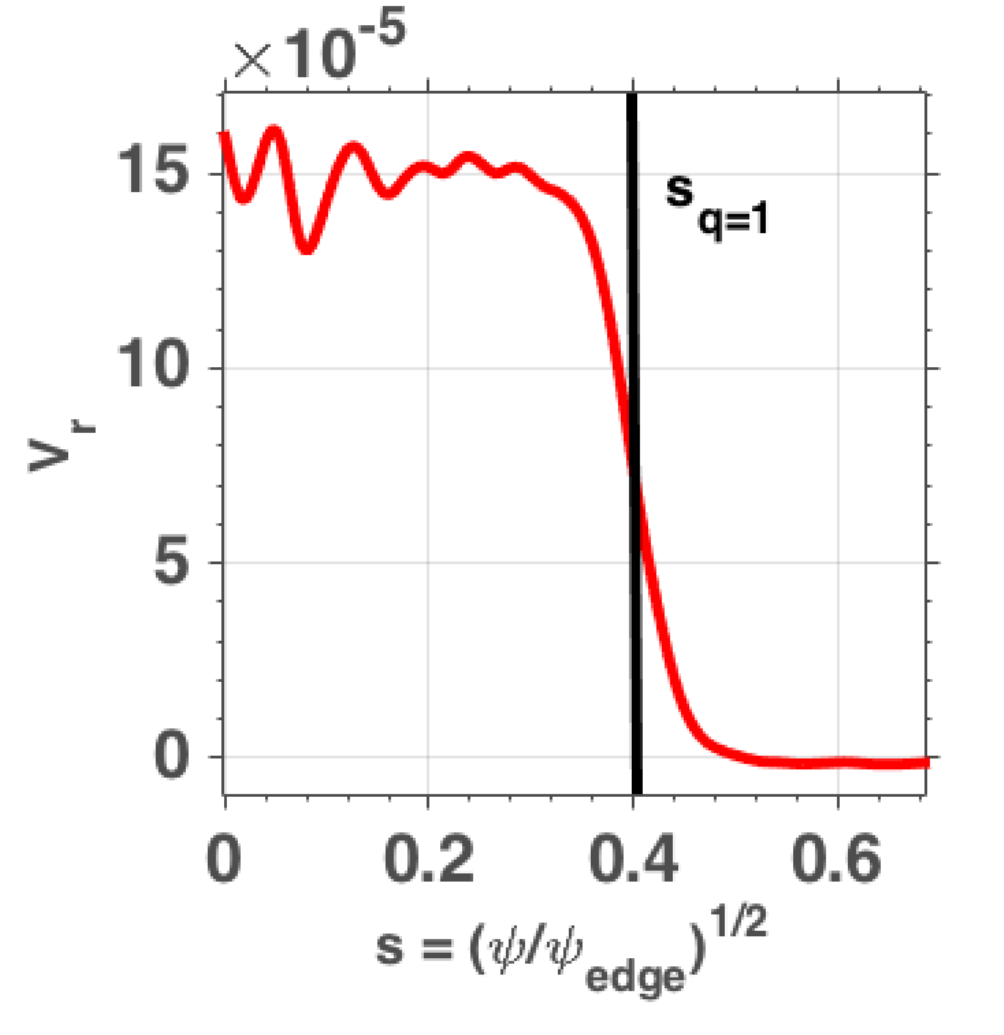}
   \caption{}
\end{subfigure}    
\caption{Radial profile of the radial MHD velocity $V_r$ obtained with XTOR-K for (a) $\beta_h/\beta_{tot} = 5.5\%$ (b) $\beta_h/\beta_{tot} = 11\%$.}
\label{xi_H}
\end{figure}
The kink branch still exists beyond $\beta_h/\beta_{tot}=5.5\%$, but has a lower growth rate than the fishbone branch. The fishbone linear model shows that fast particles fully stabilize the internal kink mode at higher kinetic beta. In XTOR-K, only the instability with the largest growth rate can be observed, which is why Figure \ref{Verif} does not show overlap between the two branches.  \\ 
\\
The agreement between the fishbone linear model and the linear simulation phases of XTOR-K is satisfactory at low kinetic densities. Both models recover the same critical kinetic beta at which the fishbone branch dominates the kink branch. On the fishbone branch, the two models begin to differ with increasing kinetic density. These differences are explained by the fact that the fishbone linear model is an asymptotic approximation. The Heaviside function used to describe the MHD displacement $\boldsymbol{\xi}$ in this model is only valid in ideal MHD with large aspect ratio. XTOR-K solves the resistive MHD equations in generalized toroidal geometry. The MHD displacement $\boldsymbol{\xi}$ that results from the dynamical evolution of the kinetic-MHD equilibrium in XTOR-K progressively departs from a Heaviside when the kinetic density is increased, as is illustrated on Figure \ref{xi_H} for $\beta_h/\beta_{tot} =$ 5.5\% and 11\%. 
\subsubsection{Quantitative agreement for the precessional resonance}
\begin{figure}[H]
\centering
\includegraphics[scale=0.3]{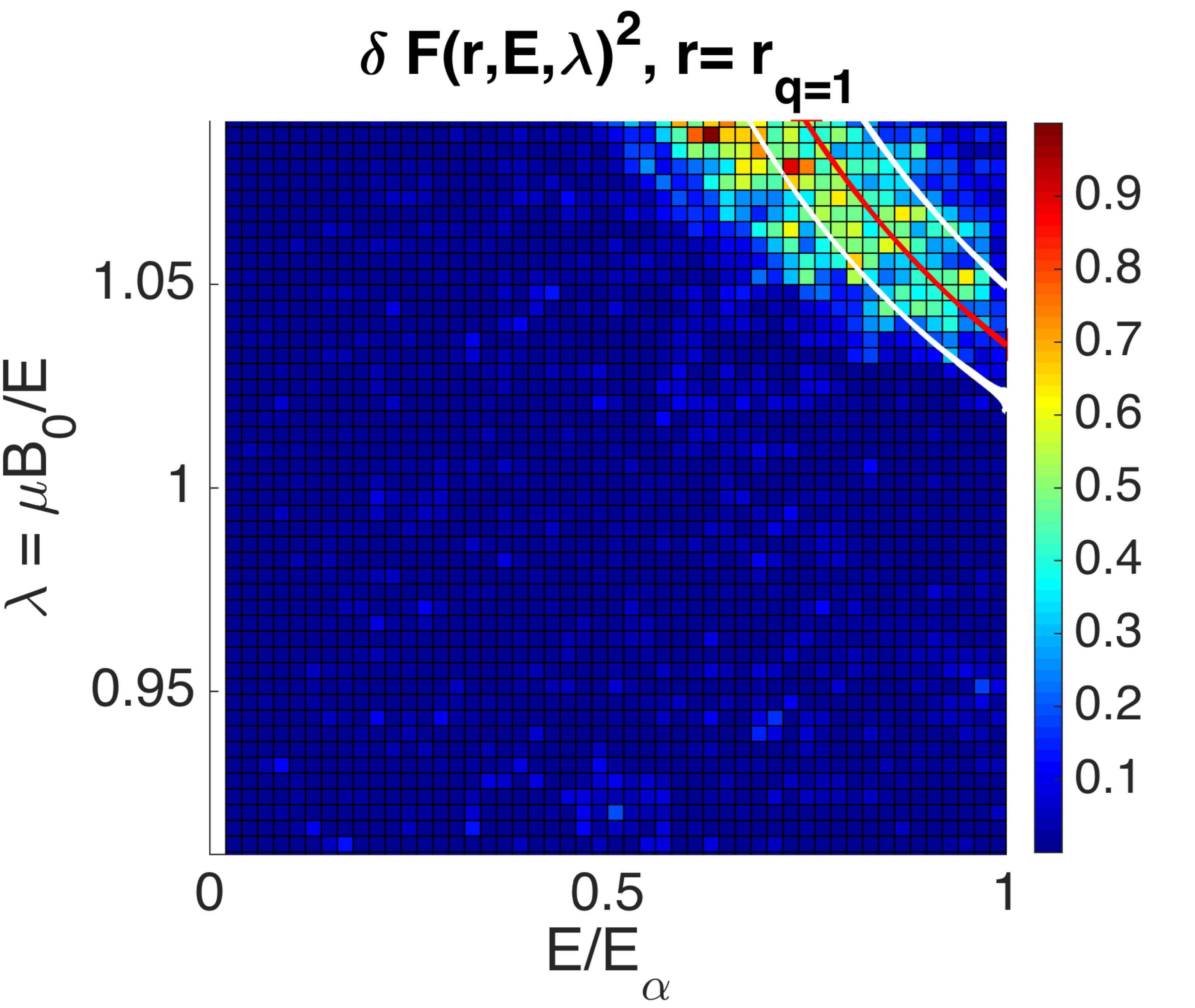}
\caption{Perturbed kinetic distribution function squared $\delta F^2$ obtained from a XTOR-K simulation in the late linear phase. $\delta F^2$ is projected on the $(E,\lambda)$ space in the radial slice $r\in [r_{q=1} -\delta r, r_{q=1} + \delta r]$. The red curve corresponds to the theoretical position of the resonance. The white curves stand for the margins of error, since $\delta F^2$ is computed on a radial slice.}
\label{res}
\end{figure}
Another comparison between the analytical model and XTOR-K is done by comparing the position in the phase space of the precessional resonance. The position of the resonance is obtained analytically by solving the equation $\omega-\omega_d(E,\lambda,\bar{r}) = 0$, with the analytical expression of $\omega_d$ derived in Annex A. This equation is solved at a fixed radial position $\bar{r}$, associated to the reference flux surface $\bar{\psi}$. The solution is a curve in the $(E,\lambda)$ phase space diagram. The resonance position is found with XTOR-K by computing the quantity $\delta F^2 = (F_h-F_{h,0})^2$ on the $(E,\lambda,\bar{r})$ grid by tri-linear interpolation, $F_{h,0}$ being the initial distribution function. During the linear phase of the fishbone instability, the perturbed kinetic distribution function $\delta F$ is expected to be maximal around the resonance position, according to Eq.(\ref{kinetic_res}). \\ 
\\
In Figure \ref{res}, both the results obtained with the linear model and XTOR-K are presented. They have been obtained at the beta ratio $\beta_h/\beta_{tot} = 8\%$ in Figure \ref{Verif}. For this simulation, the mode frequency is $\omega\tau_A = 4.10^{-3}$. The color dots in this figure correspond to the perturbed kinetic distribution function squared $\delta F^2$, computed in the $(E,\lambda)$ diagram in a radial layer in the vicinity of the $q=1$ surface $r_{q=1}$. A radial layer with finite width is necessary for a compromise between an accurate measurement and a good sampling of $\delta F^2$ on the $(E,\lambda,\bar{r})$  grid. In Figure \ref{res}, $\delta F$ corresponds to the perturbed distribution function taken at the end of the linear phase, before the fishbone mode saturates. The red curve is the solution of the resonance condition $\omega-\omega_d(E,\lambda,r_{q=1}$) in the $(E,\lambda)$ diagram. \\ 
\\
\\
Since $\delta F^2$ has been computed in XTOR-K for kinetic particles with radial positions $r\in [r_{q=1} -\delta r, r_{q=1} + \delta r]$, it is necessary to evaluate the error bars associated to this radial interval. These error bars can be provided analytically, solving $\omega - \omega(E,\lambda,r_{q=1} -\delta r)$ and $\omega - \omega(E,\lambda,r_{q=1} +\delta r)$. The solutions of these equations are illustrated respectively by the white curves under and above the red curve in Figure \ref{res}. The theoretical position of the resonance is almost identical to the resonance's position obtained from XTOR-K, the structure structure of $\delta F^2$ is aligned with the resonant domain enclosed by the white lines. \\ 
\\
Results obtained between XTOR-K and the fishbone linear model agree quantitatively regarding both the complex frequencies of the fishbone mode, and the position of precessional resonance in phase space. Therefore, a double verification has been achieved. First, these results ensure that the XTOR-K kinetic PIC module and its coupling with the fluid equations has been correctly implemented. Second, the results also show that the fishbone model used is also valid, and that the approximations used in deriving it are not too restrictive.
\subsection{Relevance of the model's specificities}
\begin{figure}[H]
\begin{subfigure}{.49\textwidth}
\centering
\includegraphics[scale=0.21]{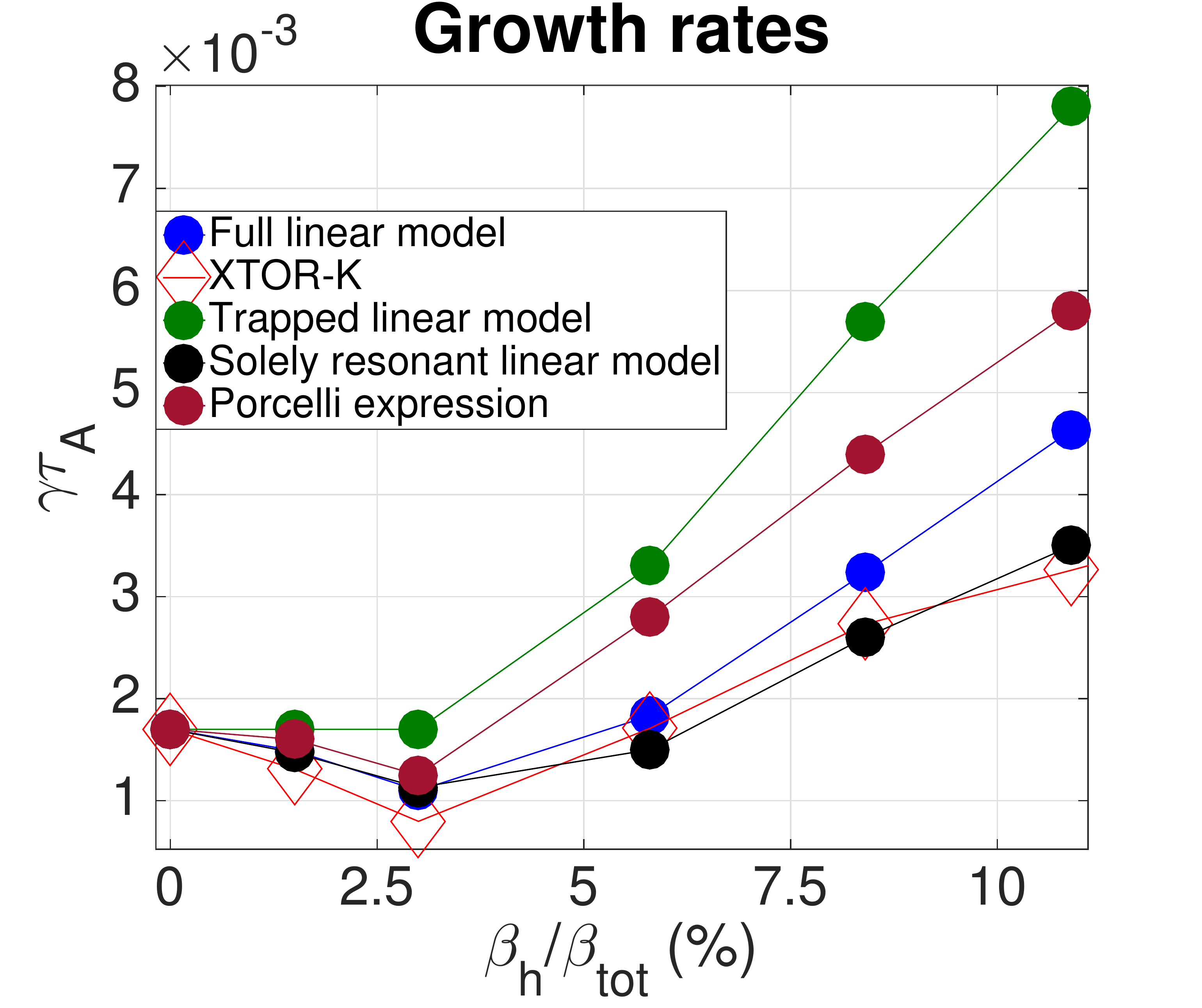}
\caption{}
\end{subfigure}
\begin{subfigure}{.49\textwidth}
\centering
\includegraphics[scale=0.21]{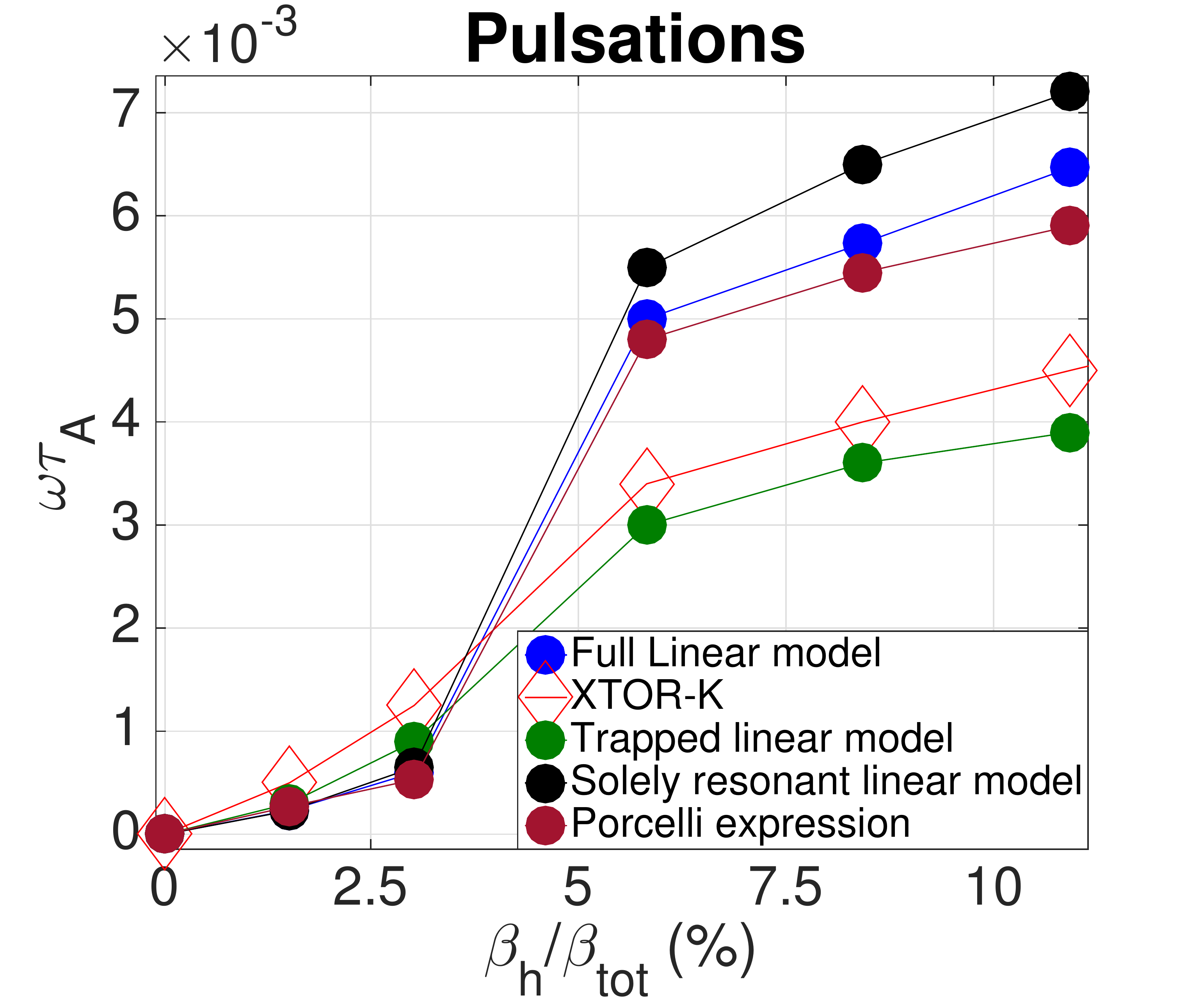}
\caption{}
\end{subfigure}
\caption{Comparisons between different theoretical linear models and XTOR-K for (a) the mode growth rates and (b) the mode frequencies. XTOR-K (Red diamonds), the full fishbone linear model (blue points), the full linear model without the resonant passing contribution (green points), the full linear model without the non-resonant contribution (black points) and  Porcelli's model expression (brown points)}
\label{comp}
\end{figure}
The specificities taken into account in the fishbone linear model are essential to provide a precise comparison with XTOR-K. Figure \ref{comp} displays the complex frequencies which are obtained with the fishbone linear model when 1) the total contribution of passing fast particles, and 2) the non-resonant kinetic contribution $\lambda_K^{int}$ to the fishbone dispersion relation is removed, respectively. Results obtained using Porcelli's expression in \cite{Porcelli1991} have also been plotted in Figure \ref{comp}. \\ 
\\
From this figure, it can be observed that without the passing contribution in the model (green points), the frequencies obtained are closer to XTOR-K values than with the complete model. However, the growth rates obtained in this limit are more than twice as high as XTOR-K growth rates. Since the linear model needs to recover precise values for the total complex frequencies, the inclusion of the contribution of passing particles is necessary for a comparison with XTOR-K. Similarly, the inclusion of the non-resonant contribution is necessary. Without this contribution (black points), the growth rates computed are closer to XTOR-K, but the frequencies are twice as large with the model. \\ 
\\
Values obtained with the complete fishbone linear model (blue points) and the Porcelli's model are also shown in Figure \ref{comp}. Complex frequencies for the Porcelli's model have been obtained by replacing the term $\sigma^2/\lambda$ in $\lambda_K^{res}$ (see Annex B), by $\lambda$ according to Eq.(11) in \cite{Porcelli1991}. The growth rates computed from this model are somewhat larger than those of the fishbone model, whereas the frequencies obtained are almost identical. 

The fishbone model derived in \cite{Brochard2018} gives the closest results with XTOR-K in the linear growth phase of the fishbones. 
\section{Linear stability of the ITER 15 MA scenario against the fishbone instability}
Now that XTOR-K and the linear fishbone model have been successfully compared, the code is used to study the linear stability of alpha fishbone modes in a regime which is not covered by the linear model : the ITER 15 MA baseline scenario. Previous linear works \cite{Fu2006}\cite{Hu2006} have examined the stability of the alpha fishbone mode on this configuration. The Kinetic-MHD equilibria used in the present work are first discussed. Then the previous linear works are detailed and compared to XTOR-K results, highlighting a disagreement between the studies for similar equilibrium parameters. Finally, XTOR-K results are presented. The fishbone thresholds are identified for the considered equilibria.
\subsection{Kinetic-MHD equilibrium for the ITER 15 MA scenario}
\begin{figure}[H]
\begin{subfigure}{.24\textwidth}
\centering
\includegraphics[scale=0.16]{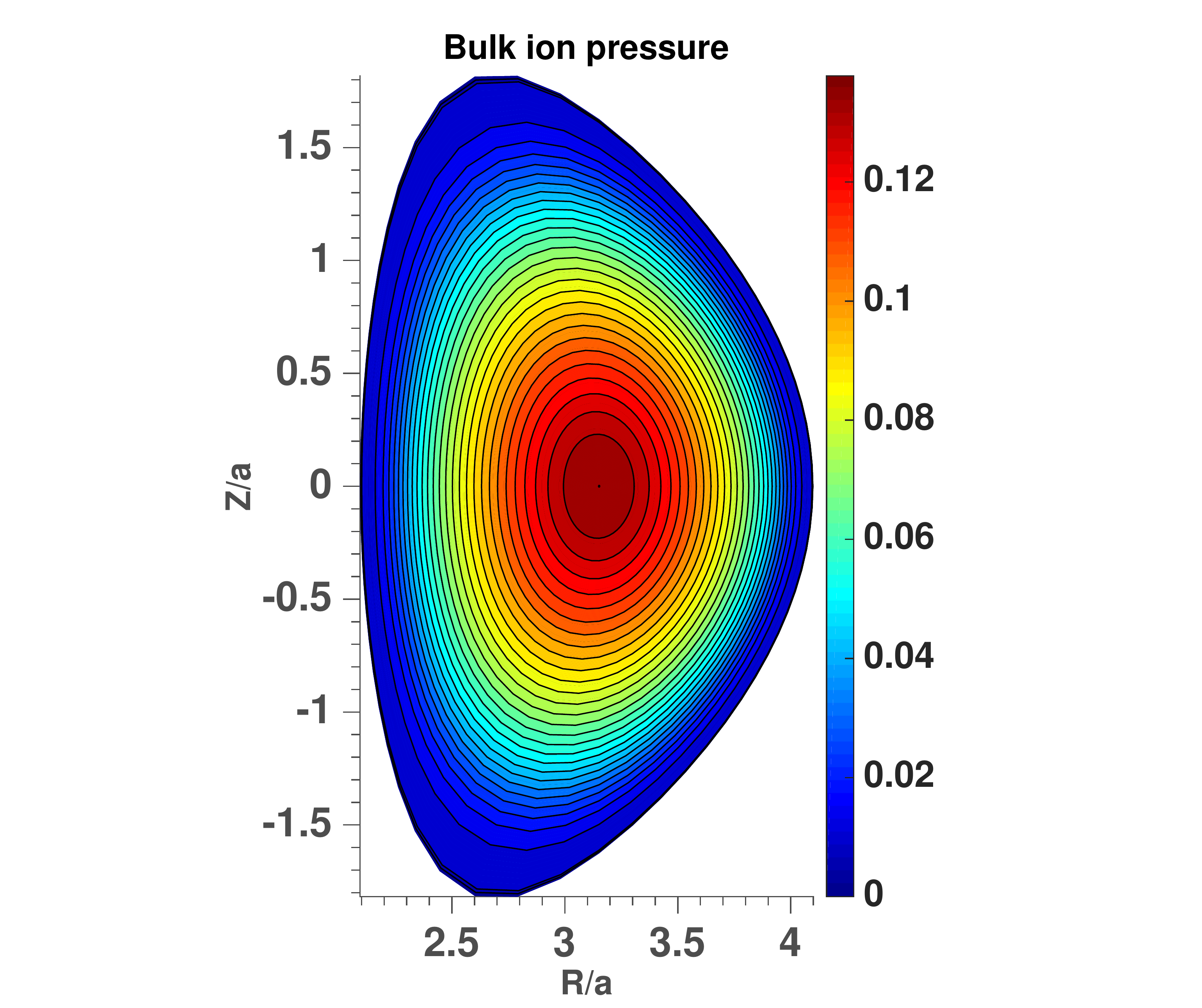}
\caption{}
\end{subfigure}
\begin{subfigure}{.24\textwidth}
\centering
\includegraphics[scale=0.16]{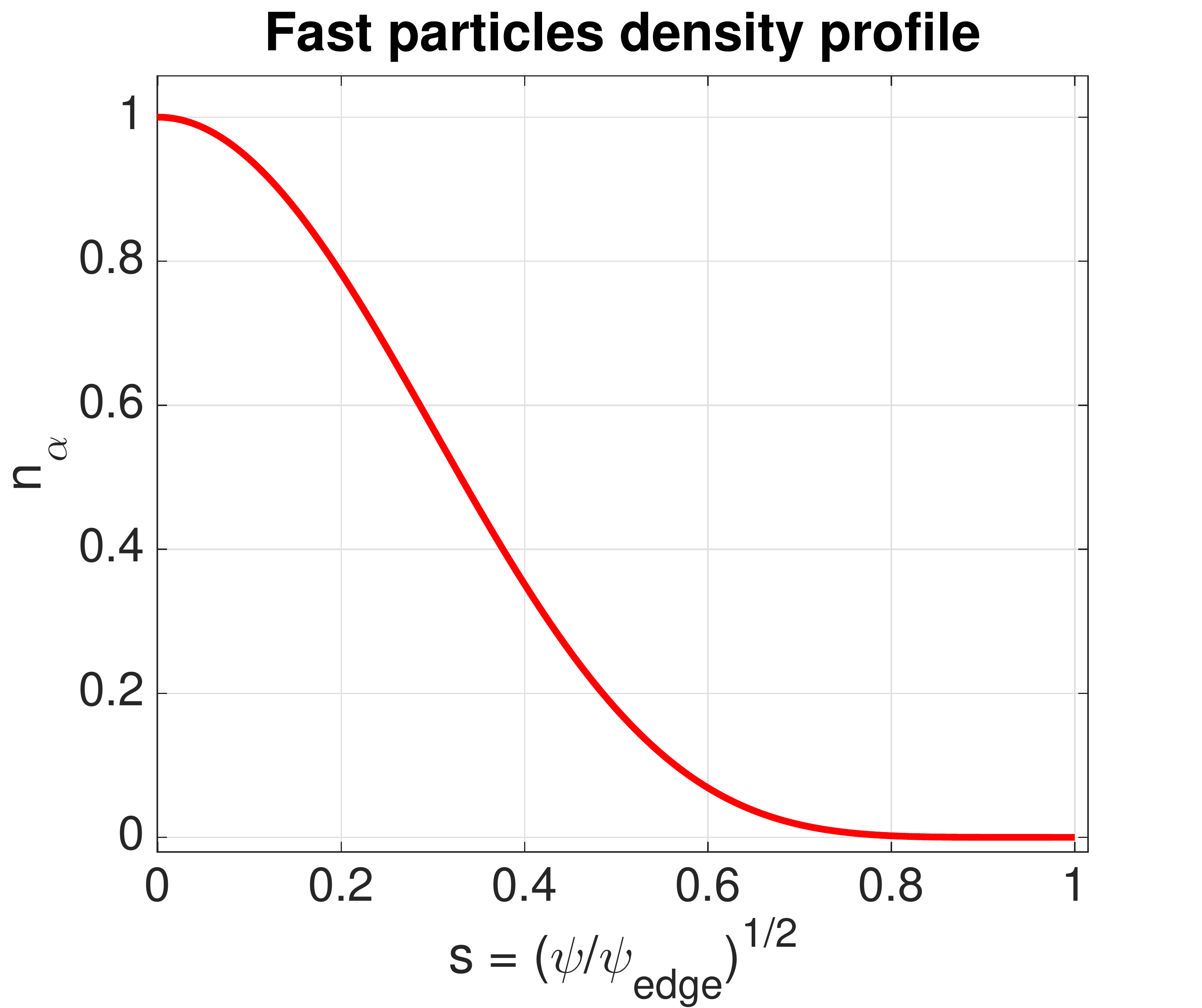}
\caption{}
\end{subfigure}
\begin{subfigure}{.24\textwidth}
\centering
\includegraphics[scale=0.16]{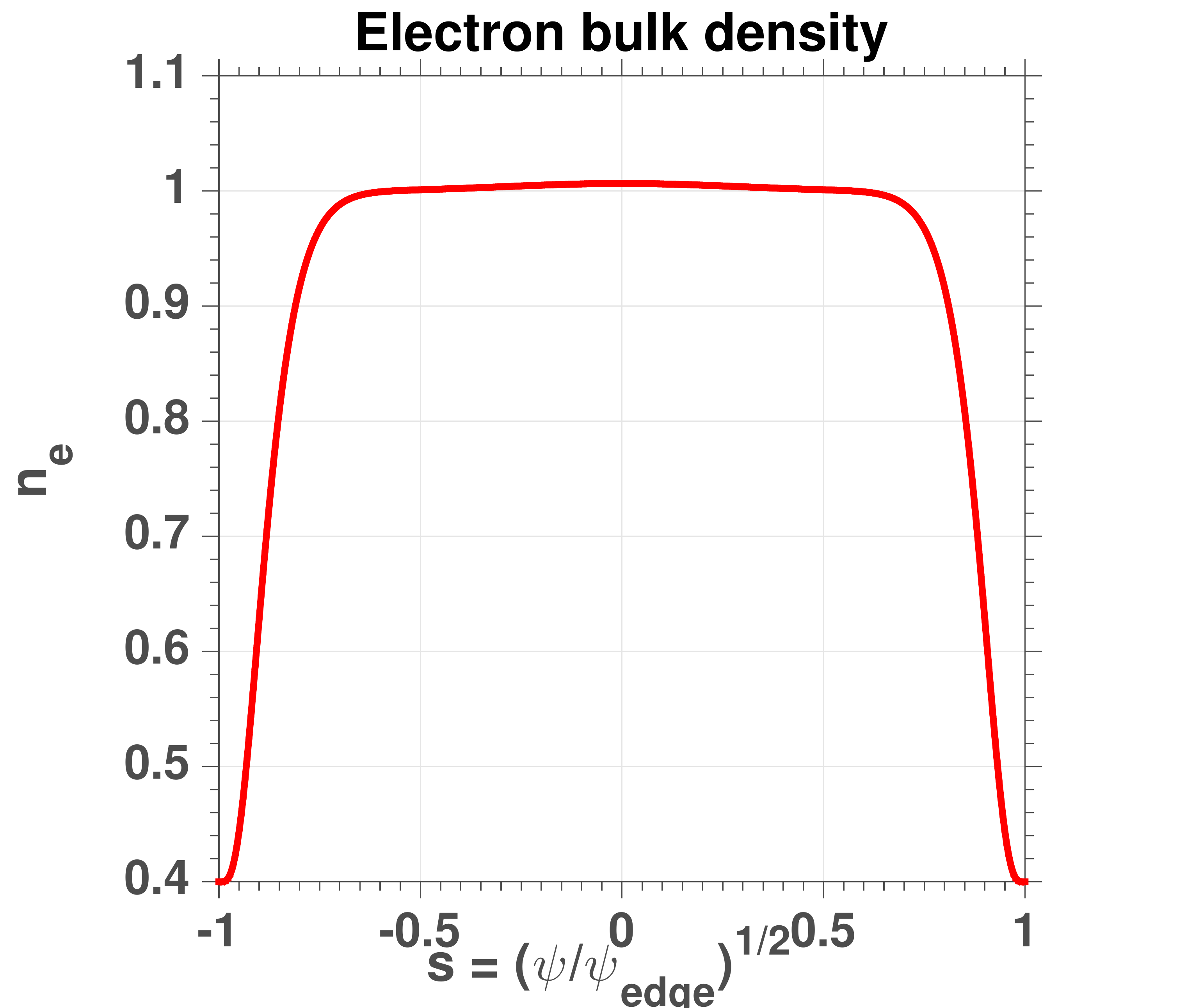}
\caption{}
\end{subfigure}
\begin{subfigure}{.24\textwidth}
\centering
\includegraphics[scale=0.16]{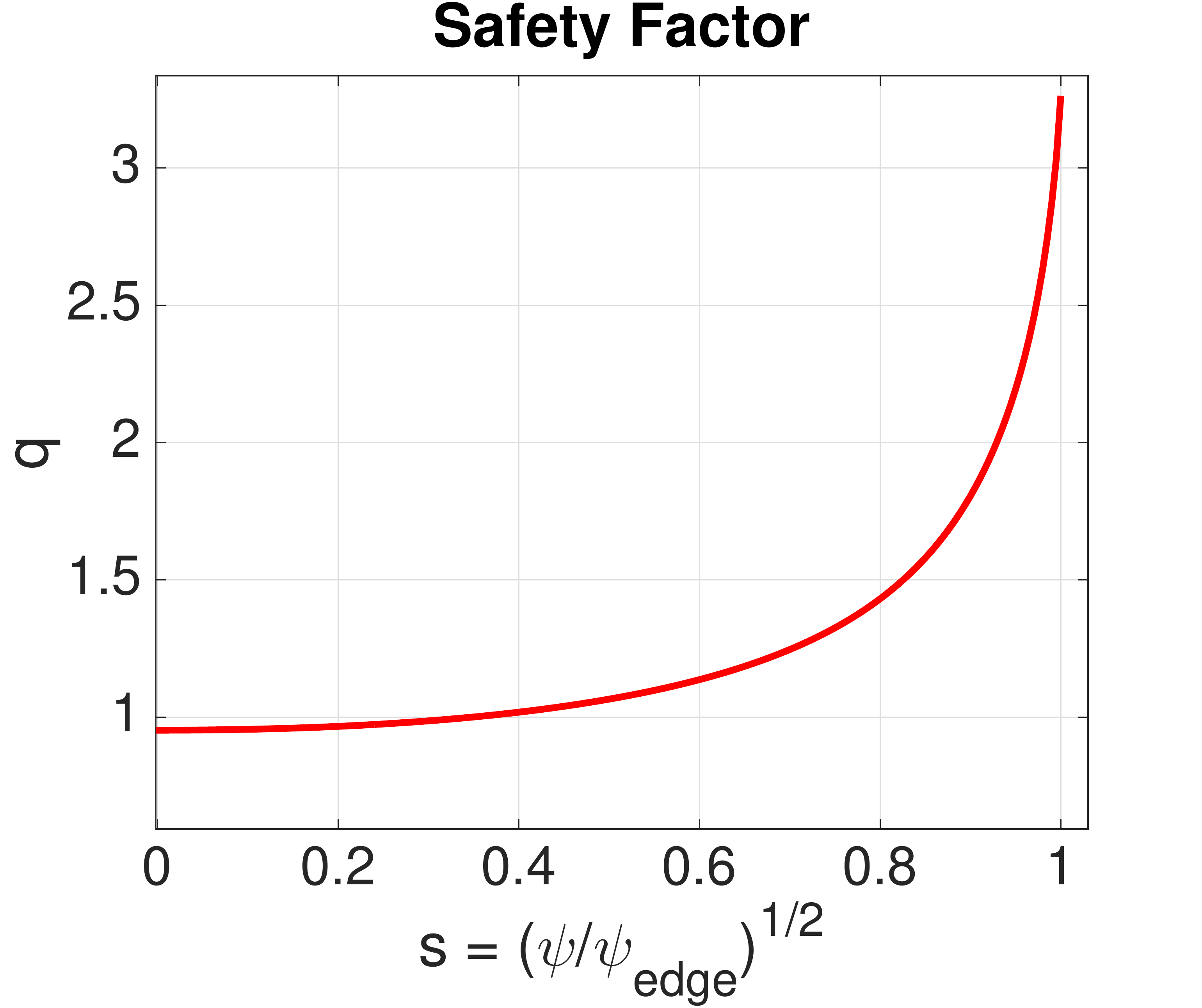}
\caption{}
\end{subfigure}
\caption{Kinetic-MHD equilibrium defined for the ITER 15 MA scenario. (a) Bulk ion pressure contour (b) Alpha particles density profile (c) Electron bulk density profile (d) Parabolic q profile}
\label{ITER_eq}
\end{figure}
The study conducted in the present work is based on profiles taken from integrated simulations performed with the code Corsica \cite{Casper2013}. Profiles have been adjusted to impose a zero pressure gradient at the plasma edge. The current profile has been modified in order to obtain parabolic $q$ profiles with on-axis value below unity using CHEASE. For the simulations, the plasma resistivity has been increased from $S = 3.10^{9}$ to $S = 1.10^{7}$ in order to resolve the mode inertial layer. $S\propto 1/\eta$ is the Lundquist number linked to the plasma resistivity.\\ \\
200 grid points are used in the radial direction. This modification of the resistivity should not affect results obtained from the fishbone simulations, since the resistivity only affects the $q=1$ layer, and the entire $q=1$ volume drives the fishbone instability. The on-axis ion/electron bulk temperatures are set at 20 keV, the bulk density at $10^{20}$ m$^{-3}$. In the different simulations performed, the only free parameters are the on-axis kinetic density, and the shape of the current profile. They are used respectively to explore the different instability branches along $\beta_{\alpha}/\beta_{tot}$, and to set the on-axis safety factor. The kinetic density profile is the same as the one used for the linear verification.  An isotropic slowing distribution function with birth energy $E_{b} = 3.5$ MeV is used to describe the alpha particles. In Figure \ref{ITER_eq}, several features of the Kinetic-MHD equilibrium are presented. 

\subsection{Previous stability studies on ITER}
In \cite{Hu2006} results have been obtained with a linear model fairly similar to our fishbone linear model. Shaped equilibria and large orbit widths are considered in this model, enabling to study the ITER configuration. Kinetic effects of the bulk plasma are also derived, to take into account the kinetic bulk ion inertia enhancement. Such bulk kinetic effects are not taken into account in XTOR-K's alpha fishbone simulations. \\ In \cite{Hu2006}, the stability region for the alpha fishbone is obtained by solving the fishbone dispersion relation at marginal stability, $\emph{i.e.}$ imposing $\gamma=0$. The stability region is computed in the diagram $[r_{q=1},\beta_{tot}]$ on Figure 5 of \cite{Hu2006}, where these two parameters vary as $r_{q=1}\in[0.3,0.5]$ and $\beta_{tot} \in [4\%,10\%]$. The beta ratio is fixed to $\beta_{\alpha}/\beta_{tot} = 7\%$. ITER relevant geometry and profiles are taken from \cite{ITERB}.
\\ \\
In \cite{Fu2006}, global hybrid simulations performed with M3D-K \cite{Park1999} are used to investigate ITER stability with respect to the alpha fishbone. One ITER relevant Kinetic-MHD linear simulation is performed, for $q_0 = 0.9$, $r_{q=1}=0.5$ and $\beta_{\alpha}/\beta_{tot} = 15\%$ (with $\beta_{tot} = 6.5\%$). In [10], Figure 3, the mode has a growth rate of $\gamma\tau_A = 6.10^{-4}$, which is 50\% less than the fluid growth rate $\gamma_{MHD}\tau_A = 1.1\times 10^{-3}$. This result disagrees with the stability region found in \cite{Hu2006}. For $r_{q=1} = 0.5$ and $\beta_{tot}=6.5\%$ in \cite{Hu2006}, the alpha fishbone mode is unstable.
\\ \\
The equilibria considered in \cite{Hu2006} can be compared to XTOR-K's equilibrium with $q_0=0.9$ and $\beta_{\alpha}/\beta_{tot}=8\%$, with $\beta_{tot} =6.23\%$. For these parameters in \cite{Hu2006}, the fishbone mode is stable, but very close to the fishbone threshold since at $r_{q=1}=0.35$, the fishbone is triggered at $\beta_{tot} = 6.6\%$. In the present work, for this set of simulations, the fishbone is triggered for $\beta_{\alpha}/\beta_{tot}\in[10\%,12\%]$ as can be observed on the red curve in Figure 12, which compares rather well with results from \cite{Hu2006} since for this beta ratio range, $\beta_{tot} \in [6.4\%,6.5\%]$. \\ 
The hybrid simulation performed in \cite{Fu2006} can be compared to one of XTOR-K simulation, with $\beta_{\alpha}/\beta_{tot} = 12\%$ and $q_0 = 0.9$. The radial position of the $q = 1$ surface is however different, with $r_{q=1} = 0.5$ in \cite{Fu2006} and $r_{q=1} = 0.35$ in the present work. The XTOR-K hybrid simulation shows that the fishbone mode is unstable with $r_{q=1}=0.35$. In [10], with $r_{q=1}=0.5$, the growth rate of the mode is smaller than the MHD growth rate with $\beta_{\alpha}=0$. Therefore it is difficult to discriminate if the mode is on the stabilized internal kink branch or the emerging fishbone branch.
 \\ \\ A parameter study with two different equilibria is done in the following section for an evaluation of the impact of the on-axis q-profile on the stability properties of the internal kink and the alpha fishbone in ITER conditions.

\subsection{Stability of alpha-fishbones in ITER 15 MA plasmas}
Two sets of simulations have been performed for this analysis. Different $q$ profiles have been used, with on-axis values of 0.9 and 0.95, same edge safety factor and same radial position for the $q=1$ surface, $r_{q=1} = 0.35$. It ensures that only the impact of the on-axis safety factor is studied when varying the $q$ profiles. The  $\beta_{\alpha}$ is increased from 0 to 12\% of the total plasma beta, which is $\beta_{tot}=5.73\%$ without alpha particles. Such a range is lower than the expected beta ratio \cite{ITERB} on ITER, where depending on the on-axis bulk temperature, $\beta_{\alpha}/\beta_{tot} \in [15\%,20\%]$. This is not restrictive since the point of these linear simulations is to find the fishbone threshold as a function of $\beta_{\alpha}/\beta_{tot}$. For both cases studied, the fishbone thresholds lie below $\beta_{\alpha}/\beta_{tot} = 12\%$. \\ 
\\
\begin{figure}[H]
\begin{subfigure}{.5\textwidth}
\centering
\includegraphics[scale=0.25]{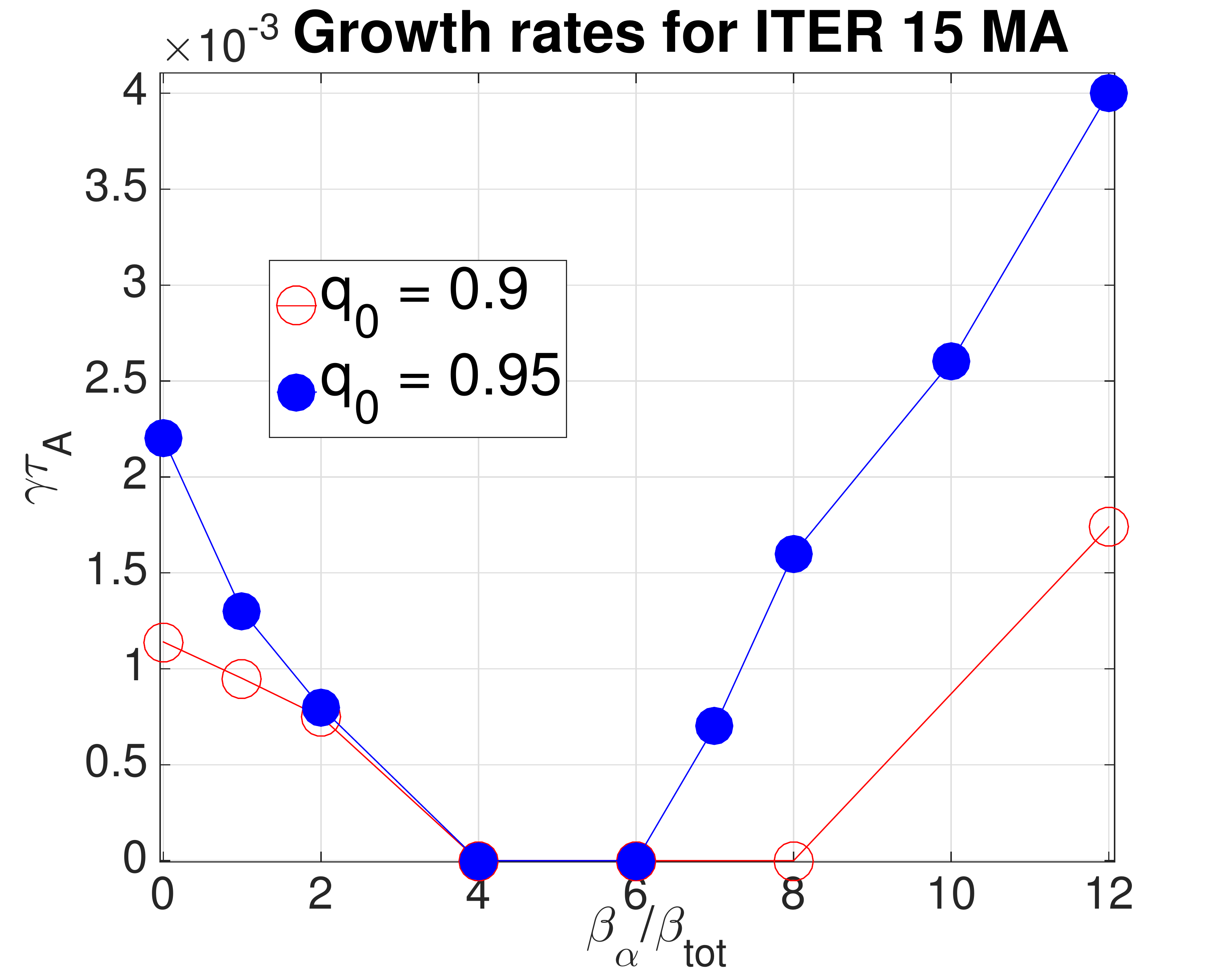}
\caption{}
\end{subfigure}
\begin{subfigure}{.5\textwidth}
\centering
\includegraphics[scale=0.25]{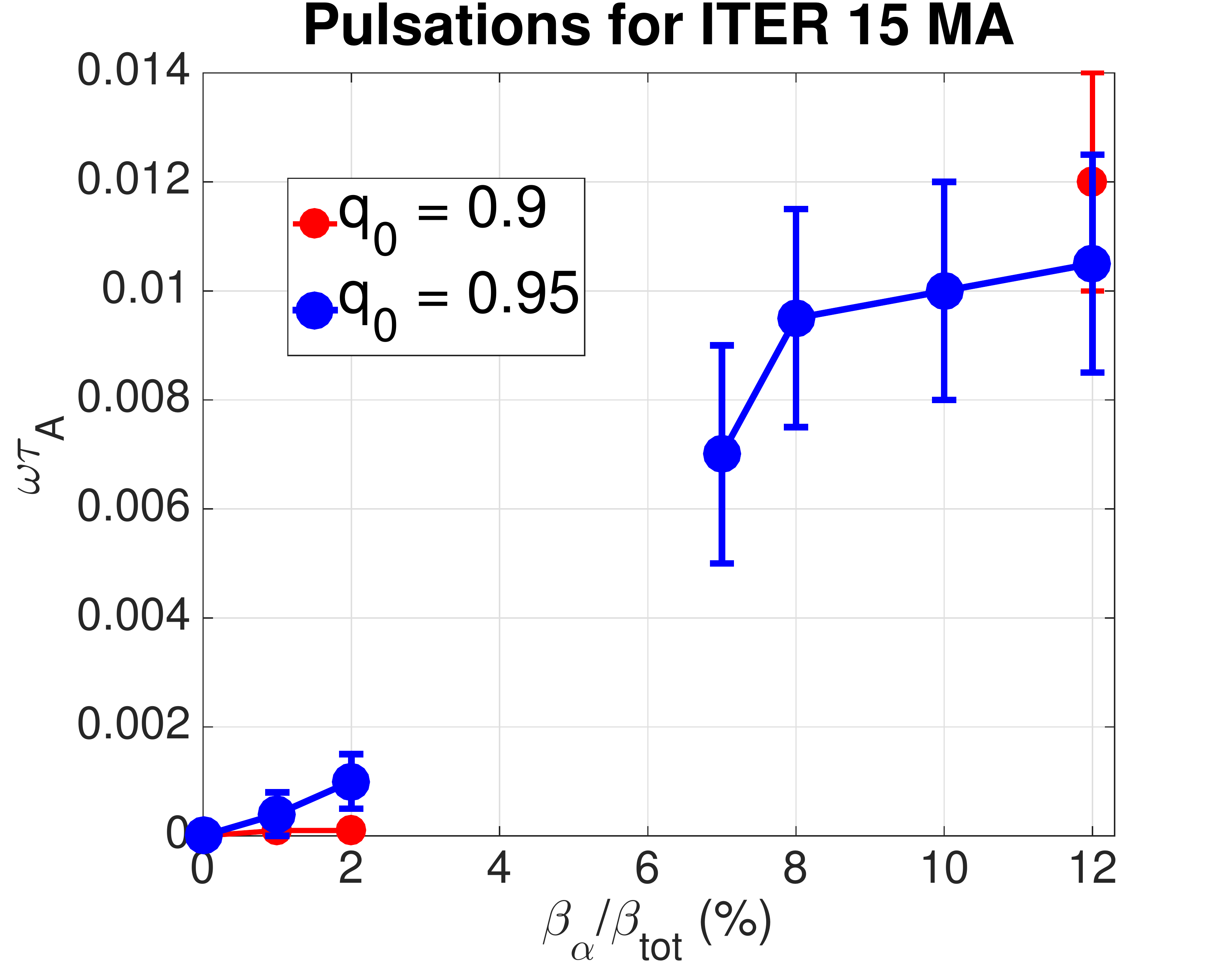}
\caption{}
\end{subfigure}
\caption{Instability growth rates (a) and frequencies (b) for the ITER 15 MA case, with in red results with $q_0=0.9$, and in blue $q_0=0.95$}
\label{ITER_ress}
\end{figure}
Similarly to the previous section, results shown in Figure \ref{ITER_ress} recover the characteristics of the interaction between fast particles and 1,1 modes. A kink and a fishbone branch appear in both cases. Points displaying null growth rates in Figure \ref{ITER_ress} (a) are cases for which the mode did not emerge from the numerical PIC noise after $4000\tau_A$. For these cases,  $\gamma\tau_A\le 6.10^{-4}$. \\ 
\\ 
These results show that the fishbone threshold is a decreasing function of the on-axis safety factor. For $q_0 = 0.95$, the threshold is located at $\beta_{\alpha}/\beta_{tot} = 6\%$, while for $q_0 = 0.9$, the fishbone branch starts around 10\%. Given the small variation applied on $q_0$, the fishbone threshold is found to be quite sensitive to the on-axis safety factor. The growth rates without alpha particles are different between the two sets of simulations.
The fluid growth rate derived in \cite{Bussac1975} scales like $1-q_0$, which explains the factor 2 of difference between these growth rates.  \\ 
\\
The error bars in Figure \ref{ITER_ress} (b) are due to the shaping of the ITER equilibrium. The mode frequency is obtained in XTOR-K by computing $\omega = \omega_{E\times B} - \omega_{lab}$. $\omega_{E\times B}$ refers to the cross field rotation of the whole plasma at $r=r_{q=1}$, and can be easily computed by projecting the MHD velocity on $n=m=0$. In principle $\omega_{E\times B}$ is a function of the radial coordinate since the MHD velocity has a radial dependency \cite{Graves2000}\cite{Graves2003}. However, for these simulations, the plasma flow is weakly sheared inside $q=1$. $\omega_{E\times B}$ can be arbitrarily defined at $r=r_{q=1}$. $\omega_{lab}$ refers to the rotation rate of the mode instability in the laboratory frame. It is computed by locating the maxima of a perturbed quantity on the flux surface $q=1$, at a given toroidal position. An error is made on this measurement since flux surfaces are not circular, the instantaneous frequency is therefore a slightly varying function of $\theta$.  \\ \\
The results obtained with XTOR-K hybrid simulations reveal that both Kinetic-MHD equilibria studied here become unstable against the fishbone mode at low kinetic beta.  For the two sets of simulations performed, the fishbone thresholds lie below the expected $\beta_{\alpha}/\beta_{tot} = 15-20 \%$ beta ratio in \cite{ITERB}, by a factor up to 3 for the set $q_0=0.95$. The alpha fishbone instability can therefore be unstable in the ITER 15 MA baseline scenario, according to the simulations presented here.
\section{Conclusion}
The Kinetic-MHD nonlinear code XTOR-K and the fishbone linear model developed in \cite{Brochard2018} have been successfully verified against each other in the linear growth regime of fishbone instabilities. For their comparison, a Kinetic-MHD equilibrium with characteristics relevant for experiments was carefully selected in order to respect the restrictive approximations of the fishbone linear model. A quantitative agreement is found between XTOR-K and the fishbone linear model for both the complex frequencies $\omega+i\gamma$, and the position in phase-space of the precessional resonance. \\ \\
A comparison was provided regarding ITER stability between the present work and previous ones \cite{Fu2006}\cite{Hu2006}. Results obtained between XTOR-K and \cite{Hu2006} agree rather well with each other concerning the fishbone threshold value. The comparison with the single simulation in \cite{Fu2006} is difficult, because the $r_{q=1}$ position used is different between the simulations ($r_{q=1} = 0.5$ in \cite{Fu2006} instead of $r_{q=1} = 0.35$ in the present work). With only one simulation in \cite{Fu2006}, and a smaller growth rate than the one of the MHD internal kink, it is difficult conclude if the mode in that work is on the stabilized kink branch or the emerging fishbone branch.. \\ \\
After the successful comparison between XTOR-K and our linear model in Section 3.2, the code has been used to determine the linear stability of the ITER 15 MA scenario against the alpha fishbone instability. For two sets of linear simulations, relevant for the ITER tokamak, the fishbone thresholds are located in the interval $\beta_{\alpha}/\beta_{tot} \in [6\%,10\%]$, whereas the beta ratio on ITER is expected to be of order $\beta_{\alpha}/\beta_{tot} \in [15\%,20\%]$. This implies that the ITER 15 MA scenario can be unstable against the alpha fishbone mode.
\\ \\
The first question in the introduction concerning the impact of the alpha fishbone on the ITER configuration, i.e. its linear stability, has been addressed. In a following paper, nonlinear simulations of the ITER 15 MA baseline scenario will be presented, in order to evaluate the amount of alpha particles transported beyond $q=1$ radius by the fishbone mode. 
\section*{Aknowledgments}
G.B, R.D and X.G would like to thank J. Graves for helpful discussions about the linear fishbone model. We are grateful to  T. Nicolas and F. Orain for commenting this manuscript.
This work has been carried out within the framework of the EUROfusion Consortium and the French Research Federation for Fusion Studies and has received funding from the Euratom research and training programme 2017-2019 under grant agreement No 633053. We benefited from HPC resources of TGCC and CINES from GENCI (projects no. 0500198 and 0510813) and the PHYMATH meso-center at Ecole Polytechnique.
\appendix
\section{Derivation of the precessional frequency for an arbitrary reference magnetic flux surface}
In this Annex, explicit derivations of the precessional drift frequency of both trapped and passing particles are presented. These derivations are performed assuming a MHD equilibrium with circular flux surfaces and low Shafranov shift. Given that for passing particles, there is no bijection between their toroidal canonical momentum $P_{\varphi}$ and the radius of their reference flux surface $\bar{r}$, the definition of their precessional drift frequency is not unique, and depends on the arbitrary choice made for the reference flux surface. First, a general derivation of $\omega_d$ will be performed without specifying $\bar{\psi}$ the reference flux surface. Then, it will be applied to two definitions, present in the literature. Analytical expressions derived in \cite{Brochard2018} Annex A will be used here.
\subsection{General expression of $\omega_d$}
Considering a general definition of the reference magnetic surface $\bar{\psi}$, as 
\begin{equation}
\bar{\psi} = \psi_0 - \frac{P_{\varphi}}{Ze}
\end{equation}
where $\psi_0$ is an arbitrary shift from the toroidal canonical momentum, the excursion from the reference magnetic surface reads
\begin{equation}
\hat{\psi} = \frac{mRv_{\parallel}}{Ze} - \psi_0
\end{equation}
The general definition for the precessional drift frequency is, with $\alpha_2$ the second angle in the angle-action formalism
\begin{equation}\label{omegad}
\omega_d = \bigg\langle \textbf{v}_d\cdot\nabla\varphi - q(\bar{\psi})\textbf{v}_d\cdot\nabla\theta + \frac{dq}{d\psi}(\bar{\psi})\hat{\psi}\frac{d\theta}{dt} \bigg \rangle_{\alpha_2}
\end{equation}
with
\begin{equation}
\textbf{v}_d = - \frac{\sigma E}{ZeB_0R_0}(\sin\theta\ \textbf{e}_r + \cos\theta \ \textbf{e}_{\theta})
\end{equation}
Therefore, the first term in Eq.(\ref{omegad}) vanishes, and the second one reads
\begin{equation}
\Big\langle-q(\bar{\psi})\textbf{v}_d\cdot\nabla\theta\Big\rangle_{\alpha_2} = \frac{q(\bar{\psi})E}{ZeB_0R_0\bar{r}} \Big[\lambda \langle \cos\theta\rangle_{\alpha_2} - \lambda\epsilon\langle\cos^2\theta\rangle_{\alpha_2}  + 4 \epsilon\lambda y^2\langle\cos\theta(1-y^{-2}\sin^2\theta/2)\rangle_{\alpha_2} \Big]
\end{equation}
In the rest of this derivation, only the lowest order in $\epsilon$ is kept. Bounce-averaged quantities for terms with higher orders in $\epsilon$ are given in Annex A in \cite{Brochard2018}. The last term in Eq.(\ref{omegad}) can be recast as
\begin{equation}
\bigg <  \frac{dq}{d\psi}(\bar{\psi})\hat{\psi}\frac{d\theta}{dt} \bigg >_{\alpha_2} = \bigg <  \frac{dq}{d\psi}(\bar{\psi})\hat{\psi}v_{\parallel}\nabla_{\parallel}\theta \bigg >_{\alpha_2}  =  \bigg <  \frac{dq}{d\psi}(\bar{\psi})\bigg[\frac{mRv_{\parallel}}{Ze} - \psi_0\bigg]v_{\parallel}\nabla_{\parallel}\theta \bigg >_{\alpha_2}
\end{equation}
Knowing that in the cylindrical limit and at leading order $\nabla_{\parallel} = \partial_{\varphi}/R_0 + R_0\partial_{\theta}/(q(\bar{\psi})R^2)$, it yields
\begin{equation}\label{ltm}
\bigg <  \frac{dq}{d\psi}(\bar{\psi})\hat{\psi}\frac{d\theta}{dt} \bigg >_{\alpha_2} = 4 \frac{dq}{d\psi}(\bar{\psi})\frac{\bar{r}\lambda E y^2}{q(\bar{\psi})ZeR_0} \times  \Big[ \frac{m}{4E\lambda y^2\epsilon}\langle v^2_{\parallel}\rangle_{\alpha_2} - \frac{2Ze}{R_0m}\Big(\frac{m}{E\epsilon\lambda}\Big)^{1/2} \langle\psi_0\rangle_{\alpha_2} \Big]
\end{equation}
The arbitrary reference flux surface $\bar{\psi}$ being linked to its reference radius as $\bar{\psi} = B_0\bar{r}^2/2q(\bar{r})$ without loss of generality, the derivative along $\psi$ can be recast as
\begin{equation}
\frac{dq}{d\psi}(\bar{\psi}) = \frac{s(\bar{r})}{\bar{r}^2B_0}, \  \  \  \  \  \ s(\bar{r}) = \bar{r}\frac{dq}{dr}(\bar{r})/q(\bar{r})
\end{equation}
Eq.(\ref{ltm}) can then be re-expressed as
\begin{equation}
\bigg <  \frac{dq}{d\psi}(\bar{\psi})\hat{\psi}\frac{d\theta}{dt} \bigg >_{\alpha_2} = \frac{q(\bar{r})\lambda E}{ZeB_0\bar{r}R_0}4s(\bar{r})y^2\times \Big[ \frac{m}{4E\lambda y^2\epsilon}\langle v^2_{\parallel}\rangle_{\alpha_2} - \frac{2Ze}{R_0m}\Big(\frac{m}{E\epsilon\lambda}\Big)^{1/2} \langle\psi_0\rangle_{\alpha_2} \Big]
\end{equation}
The precessional drift frequency for an arbitrary reference flux surface is then, at lowest order in $\epsilon$
\begin{equation}
\omega_d = \frac{q(\bar{r})\lambda E}{ZeB_0\bar{r}R_0}\bigg[\langle\cos\theta\rangle_{\alpha_2}  + 4s(\bar{r})y^2\times \bigg( \frac{m}{4E\lambda y^2\epsilon}\langle v^2_{\parallel}\rangle_{\alpha_2} - \frac{2Ze}{R_0m}\Big(\frac{m}{E\epsilon\lambda}\Big)^{1/2} \langle\psi_0\rangle_{\alpha_2} \bigg)\bigg]
\end{equation}
\subsection{Explicit expressions}
\subsubsection{Trapped particles}
For trapped particles, the choice for $\psi_0$ does not matter since their reference flux surface is an invariant of motion. The choice of the reference flux surface is then unique and intersects the banana turning points in the poloidal plane as $\bar{\psi} = \psi = -P_{\varphi}/Ze$. The explicit expression for the precessional frequency is then, using Annex A
\begin{equation}\label{omegadt}
\omega_d(\bar{r},\lambda,E) = \frac{q(\bar{r})\lambda E}{ZeB_0\bar{r}R_0}I_{d,t}(\bar{r},\lambda)
\end{equation}
\begin{equation}
I_{d,t} =\bigg[2\frac{\mathbb{E}(y^2)}{\mathbb{K}(y^2)} -1 + 4s(\bar{r})\Big(\frac{\mathbb{E}(y^2)}{\mathbb{K}(y^2)} + y^2 - 1 \Big) \bigg]
\end{equation}
\subsubsection{Passing particles}
Two definitions of the reference flux surface for passing particles are present in the literature. In \cite{Nguyen2009}, this surface is taken to be directly proportional to the canonical toroidal momentum, $\bar{\psi} = -P_{\varphi}/Ze$, where in this case $\psi_0 = 0$. $\bar{\psi}$ therefore corresponds to the flux surface of the trapped particles banana tips. This choice enables to simplify significantly the derivations using the angle-action formalism, and is closed to the intrinsic reference flux surface for passing particles near the passing-trapped frontier. The explicit expression for $\omega_d$ in this case is
\begin{equation}
\omega_d(\bar{r},\lambda,E) = \frac{q(\bar{r})\lambda E}{ZeB_0\bar{r}R_0}I_{d,p}(\bar{r},\lambda)
\end{equation}\label{omegadno}
\begin{equation}
I_{d,p} = \bigg[2y^2\bigg(\frac{\mathbb{E}(1/y^2)}{\mathbb{K}(1/y^2)} - 1\bigg) + 1 + 4s(\bar{r})y^2\frac{\mathbb{E}(1/y^2)}{\mathbb{K}(1/y^2)}   \bigg]
\end{equation}
It is noted that this expression is identical in \cite{Nguyen2009}. However, this definition is not the most practical. \\
\\
Indeed, for energetic particles, the term $mRv_{\varphi}$ is of the same order of magnitude as $Ze\psi$. Therefore, passing particles orbits can be quite distant from the latter definition of $\bar{\psi}$. If one wishes to compare theoretical values for $\omega_d$ with ones obtained from orbit codes, as it is done in section 3 with XTOR-K, it prevents a precise comparison. Therefore, a wiser choice is to take the time averaged particle flux surface as reference, which is equivalent to take its average value along $\alpha_2$
\begin{equation}
\bar{\psi} = \langle{\psi}\rangle_t = \frac{1}{Ze}[m\langle Rv_{\parallel}\rangle_{\alpha_2} - P_{\varphi}], \ \ \ \ \  \hat{\psi} = \frac{m}{Ze}[Rv_{\parallel} - \langle Rv_{\parallel}\rangle_{\alpha_2}]
\end{equation}
Such a choice is also made in \cite{Graves2013}\cite{M}\cite{Zonca2007}. It implies that $\psi_0 = m\langle R v_{\parallel}\rangle_{\alpha_2}/Ze$, the explicit expression for $\omega_d$ then reads
\begin{equation}\label{omegadp}
\omega_d(\bar{r},\lambda,E) = \frac{q(\bar{r})\lambda E}{ZeB_0\bar{r}R_0}\bigg[2y^2\bigg(\frac{\mathbb{E}(1/y^2)}{\mathbb{K}(1/y^2)} - 1\bigg) + 1 + 4s(\bar{r})y^2\bigg(\frac{\mathbb{E}(1/y^2)}{\mathbb{K}(1/y^2)} - \Big(\frac{\pi}{2\mathbb{K}(1/y^2)}\Big)^2\bigg)\bigg]
\end{equation}
This expression agrees with \cite{M} and is close to \cite{Zonca2007}, up to the term $(\pi/2\mathbb{K}(1/y^2))^2$, replaced by $(\pi/2\mathbb{K}(1/y^2))\sqrt{1-y^{-2}}$
\section{Analytical expression and integration of $\lambda_K^{res}$}
\subsection{General derivation}
A correction to \cite{Brochard2018} is brought here regarding the analytical expressions of $\lambda_K^{res}$'s integrand in invariants space. A factor $\lambda\sigma$ in the pitch angle integral of \cite{Brochard2018} Eq.(53) was mistakenly used instead of $\sigma^2/\lambda$. The complete expression for $\lambda_K^{res}$ can be split in two parts as $\lambda_K^{res} = \lambda_K^{res,\omega_*} + \lambda_K^{res,\omega} $, where $\lambda_K^{res,\omega_*}$ corresponds to the term proportional to $\omega_*$, and $\lambda_K^{res,\omega}$ the one proportional to $\omega$. Following the notations used in \cite{Brochard2018}, the complete equations for these terms are 
\begin{equation}
\lambda_K^{res,\omega_*} = \frac{3\pi^2\epsilon_0E_b}{2s_0 r_0 B_{p,0}^2\ln[1+(v_b/v_c)^3]}\sum_{\sigma_{\parallel} = \pm 1}\times \int_0^{r_0}\frac{dn_{\alpha}}{d\bar{r}} \bar{r}d\bar{r}  \int_{0}^{(1-\epsilon)^{-1}}d\lambda\frac{\sigma^2I_bI_q^2}{\lambda I_d}I_{res,1}
\end{equation}
with
\begin{equation}
I_{res,1} = \frac{4}{v_+ - v_-}\bigg[v_+\int_0^1d\hat{v}\frac{\hat{v}^5}{(\hat{v}^3 + \hat{v}_c^3)(\hat{v}-v_+)} - v_-\int_0^1d\hat{v}\frac{\hat{v}^5}{(\hat{v}^3 + \hat{v}_c^3)(\hat{v}-v_-)}\bigg]
\end{equation}
\begin{equation}
\lambda_K^{res,\omega} = \frac{3\pi^2\epsilon_0^2\hat{\omega}E_{b}}{2s_0 B_{p,0}^2\ln[1+(v_{\alpha}/v_c)^3]} \sum_{\sigma_{\parallel} = \pm 1}\times  \int_0^{r_0}\frac{x^2n_{\alpha}}{q}d\bar{r} \int_{0}^{(1-\epsilon)^{-1}}d\lambda \frac{\sigma^2I_bI_q^2}{\lambda I_d}I_{res,2}
\end{equation}
\begin{equation}
I_{res,2} = \frac{1}{(1+\hat{v}_c^3)(1+v_1+v_2)} + \frac{3}{2(v_+-v_-)} \times  \bigg[v_+\int_0^1d\hat{v}\frac{\hat{v}^3}{(\hat{v}^3 + \hat{v}_c^3)(\hat{v}-v_+)} - v_-\int_0^1d\hat{v}\frac{\hat{v}^3}{(\hat{v}^3 + \hat{v}_c^3)(\hat{v}-v_-)}\bigg]
\end{equation}
The resonant integral can be computed either analytically or numerically in the fishbone linear model. An analytical expression for these integrals can be obtained provided that the birth velocity/critical velocity ratio respects either $v_c/v_b \ll 1$ or $v_c/v_b\gg1$. If an ordering for this ratio cannot be obtained in general for most $\hat{v}$, a numerical scheme called the collocation method has to be used. \\ \\ In cases where an ordering can be obtained, the following expression enables to explicit $I_{res,1}$ and $I_{res,2}$
\begin{equation}
\int_0^1\frac{v^n}{v-v_0} = \sum_{m=1}^{n-1}v_0^m\int_0^1v^{n-m}dv + v_0^n\ln\bigg[1-\frac{1}{v_0}\bigg]
\end{equation}
\subsection{$v_c/v_b \ll \hat{v}$}
When $v_c/v_b \ll \hat{v}$, $I_{res,1}$ and $I_{res,2}$ can be recast as
\begin{equation}
I_{res,1} = \frac{4}{v_+ - v_-}\bigg[v_+\int_0^1d\hat{v}\frac{\hat{v}^2}{\hat{v}-v_+}  - v_-\int_0^1d\hat{v}\frac{\hat{v}^2}{\hat{v}-v_-}\bigg]
\end{equation}
\begin{equation}
I_{res,2} = \frac{1}{1+v_1+v_2} + \frac{3}{2(v_+-v_-)}\bigg[v_+\int_0^1d\hat{v}\frac{1}{\hat{v}-v_+}  - v_-\int_0^1d\hat{v}\frac{1}{\hat{v}-v_-}\bigg]
\end{equation}
\subsection{$v_c/v_b \gg \hat{v}$}
When $v_c/v_b \ll \hat{v}$, $I_{res,1}$ and $I_{res,2}$ can be recast as
\begin{equation}
I_{res,1} = \frac{4}{\hat{v_c}^3(v_+ - v_-)}\bigg[v_+\int_0^1d\hat{v}\frac{\hat{v}^5}{\hat{v}-v_+} - v_-\int_0^1d\hat{v}\frac{\hat{v}^5}{\hat{v}-v_-}\bigg]
\end{equation}
\begin{equation}
I_{res,2} = \frac{1}{\hat{v}_c^3}\bigg(\frac{1}{1+v_1+v_2} + \frac{3}{2(v_+-v_-)}\times \bigg[v_+\int_0^1d\hat{v}\frac{\hat{v}^3}{\hat{v}-v_+} - v_-\int_0^1d\hat{v}\frac{\hat{v}^3}{\hat{v}-v_-}\bigg]\bigg)
\end{equation}
\subsection{The collocation method}
When an ordering between $\hat{v}_c$ and $\hat{v}$ cannot be found for most $\hat{v}\in[0,1]$, the collocation method is used. The method aims at computing the following resonant integral
\begin{equation}
K = \int_{-\infty}^{+\infty} dv\frac{g(v)}{v-v_0}
\end{equation}
In order to compute $I_{res,1}$ and $I_{res,2}$, the function $g$ is identified as $g(\hat{v}) = \hat{v}^n/(\hat{v}^3+\hat{v}_c^3)$, with $n=3$ or $5$. The collocation method consists in computing $K$ on uniformly spaced grid such as $v_0 = k\Delta v$, with $\Delta v$ the length between two grid points and $k\in[0,N]$.
\\ \\ On that grid, $g$ is approximated as
\begin{equation}
g(v) = \sum_jg_jh_j(v)
\end{equation}
with $h_i(v) = 0$ when $|v-v_j|>\Delta v$, and
\begin{equation}
h_j(v) = 1-\frac{|v-v_j|}{\Delta v}
\end{equation}
otherwise. The resonant integral $K$ can then be expressed as
\begin{equation}
K = \sum_{j=-N}^{N}g_j\kappa_{j,k}
\end{equation}
with the kernel
\begin{equation}
\kappa_{j,k} = \int_{-1}^{1}dx\frac{1-|x|}{x+j-k}
\end{equation}
The kernel can be computed analytically. When $j-k\ne 0$ and $j-k\ne\pm 1$
\begin{equation}
\kappa_{j,k} = \ln\bigg[\frac{j-k+1}{j-k-1}\bigg] - (j-k)\ln\bigg[\frac{(j-k)^2}{(j-k)^2-1}\bigg]
\end{equation}
The singularity is handled by taking $\kappa_{j,k} = \pm 2\ln(2)$ when $j-k=\pm1$, and $\kappa_{j,k} = i\pi$ when $j-k=0$.
\bibliography{thesis_lib}{}
\end{document}